\documentclass[aps,twocolumn,amsfonts,amsmath,amssymb,showkeys,nobibnotes,nofootinbib,floatfix,a4paper]{revtex4-1}

\usepackage{graphicx}  
\usepackage{amsfonts} 
\usepackage{amsmath} 
\usepackage{amssymb} 
\usepackage{hyperref}
\usepackage{cleveref}
\usepackage{enumerate}

\begin{document}

\title{Elimination of systemic risk in financial networks by means of a systemic risk transaction tax} 

\author{Sebastian Poledna$^{1}$}

\author{Stefan Thurner$^{1,2,3}$}
\email{stefan.thurner@meduniwien.ac.at}

\affiliation{
$^1$Section for Science of Complex Systems; Medical University of Vienna; Spitalgasse 23; A-1090; Austria\\
$^2$IIASA, Schlossplatz 1, A-2361 Laxenburg; Austria\\
$^3$Santa Fe Institute; 1399 Hyde Park Road; Santa Fe; NM 87501; USA}

\begin{abstract}
	Financial markets are exposed to systemic risk (SR), the risk that a major fraction of the system ceases to function, and collapses. It has recently become possible to quantify SR in terms of underlying financial networks where nodes represent financial institutions, and links capture the size and maturity of assets (loans), liabilities, and other obligations, such as derivatives. We demonstrate that it is possible to quantify the share of SR that individual liabilities within a financial network contribute to the overall SR. We use empirical data of nationwide interbank liabilities to show that the marginal contribution to overall SR of liabilities for a given size varies by a factor of a thousand. We propose a tax on individual transactions that is proportional to their marginal contribution to overall SR. If a transaction does not increase SR it is tax-free. With an agent-based model (CRISIS macro-financial model) we demonstrate that the proposed ``Systemic Risk Tax'' (SRT) leads to a self-organised restructuring of financial networks that are practically free of SR. The SRT can be seen as an insurance for the public against costs arising from cascading failure. ABM predictions are shown to be in remarkable agreement with the empirical data and can be used to understand the relation of credit risk and SR. 
\end{abstract}


\maketitle 

\section{Introduction} \label{intro} 
Failure to manage systemic risk (SR) has turned out to be extremely costly for society. The financial crisis of 2007-2008 and its consequences demonstrated the importance of reducing it. The threat of collapse of large parts of the financial system has forced national governments to bailout hundreds of banks \citep{ecobailout}. As a result, one observed falling global stock and real estate markets \citep{ecoresession}, a severe and global credit crunch \citep{ecocrunch}, skyrocketing and prolonged unemployment rates, and several Western governments at the verge of bankruptcy. Bank bailouts have caused dangerously high levels of sovereign debt around the world, and it has become necessary to find alternatives to finance bailouts \citep{Klimek:2014aa}. The International Monetary Fund has proposed a tax on banks, called the ``financial stability contribution'' (FSC), i.e. a contribution of the financial sector to the public costs of the financial crisis, which is used to create reserves for future crises. Bank taxes have been proposed in many countries around the world, e.g. the ``Financial Crisis Responsibility Fee'' in the US. In several European countries, including Germany and Austria, bank taxes are currently in force. The European Commission has proposed an EU-wide bank tax under the ``Single Resolution Mechanism''. In addition to bank taxes, a financial transaction tax (FTT) is being considered by many countries. A FTT is not a tax on financial institutions \emph{per se}, but a levy placed on specific types of financial transactions. Its main purpose, besides generating revenue for governments, is to curb the volatility of financial markets \citep{Tobin:1978aa,Summers:1989aa}. Related empirical studies are generally inconclusive, and a causal relation between volatility and FTTs remains ambiguous \citep{McCulloch:2011aa,Matheson:2012aa}. In response to the financial crisis of 2007-2008, a consensus on the need for new financial regulation is emerging \citep{Aikman:2013aa}. New financial regulation must be designed to mitigate the risk of the financial system as a whole. This approach to financial regulation is known as ``macroprudential regulation'', and is currently being put in place around the globe \citep{Aikman:2013aa,Bank-of-England:2011aa,Bank-of-England:2013aa}. The Basel III framework recognises systemically important financial institutions (SIFI) and recommends increased capital requirements for them -- the so called ``SIFI surcharges'' \citep{BIS:2010aa,Georg:2011aa}. Basel III further introduces the idea of ``counter-cyclical buffers'' that allow regulators to increase capital requirements during periods of high credit growth. No matter how well-intended these developments might be, they miss the central point about the nature of SR, and therefore may not be suitable to improve the stability of the financial system in a sustainable way. SR is closely related to the \emph{network structure} of financial assets and liabilities in a financial system. Management of SR is essentially a matter of restructuring financial networks in such a way that the probability of cascading failure is reduced, or ideally eliminated.

Credit risk is the risk that a borrower will default on a specific debt by failing to make the full pre-specified repayments. It is usually seen as a risk that emerges between two counterparties once they have engaged in a financial transaction. The lender is the sole bearer of credit risk and accounts for the likelihood of failed repayments by demanding a risk premium. Lenders usually charge higher interest rates to borrowers that are more likely to default (risk-based pricing). Credit risk is relatively well-understood, and can be mitigated through a number of methods and techniques \citep{Duffie:2012aa}. The Basel Accords provide an extensive framework, dealing foremost with the mitigation of credit risk \citep{BIS:1988aa,BIS:2006aa,Balin:2008aa}. When two counterparties are part of a financial system, for example as nodes in a financial network, the situation changes, and their transaction may affect the financial system as a whole. The lender is no more the sole bearer of credit risk, nor does credit risk depend on the financial conditions of the borrower alone. The impact of a default of the borrower is no longer limited to the lender, but may affect other creditors of the lender, which in turn may affect their creditors. Similarly, the lender is not only vulnerable to a default of the borrower but also to defaults of all debtors of that borrower, as well as their debtors. In other words, in financial networks credit risk is no longer limited to two counterparties, but becomes \emph{systemic}.

SR is the risk that the financial system as a whole, or a large fraction of it, can no longer perform its function as a credit provider and collapses. In a narrow sense, SR is the notion of contagion or impact from the failure of a financial institution, or group of institutions, on the financial system and the wider economy \citep{De-Bandt:2000aa,BIS:2010aa}. It is a result of the interconnected nature of financial transactions, and claims or liabilities in the financial system. It unfolds as secondary cascades of credit defaults, triggered by credit defaults between individual counterparties \citep{Eisenberg:2001aa}. These cascades can potentially wipe out the financial system by a de-leveraging cascade \citep{Minsky:1992aa,Fostel:2008aa,Geanakoplos:2010aa,Adrian:2008aa,Brunnermeier:2009aa,Thurner:2012aa,Caccioli:2012aa,Poledna:2014ab,Aymanns:2014aa,Caccioli:2015aa}. It is obvious that lenders have a strong incentive to mitigate credit risk. In the case of SR the situation is less clear as SR involves externalities, i.e. financial institutions manage their own risks but do \emph{not} consider their impact on the system as a whole \citep{Acharya:2009aa}. In fact, funding costs for large financial institutions are lowered due to a market expectation that the state will bailout banks that are deemed to be systemically important \citep{Davies:2014aa}. Unless financial institutions are required to internalise costs of SR, institutions will have little incentive to minimise risks that are borne by the general public \citep{Acharya:2010aa}. Management of SR is, therefore, foremost in the public interest. 

SR is a network externality resulting from contagion effects \citep{Acemoglu:2013aa}. To cope with this externality, governments can use two main policy instruments: taxation or regulation \citep{Masciandaro:2013aa}. Taxation is aimed at reducing the gap between public and private costs of SR, while financial regulations impose direct restrictions and requirements on financial institutions. In general, taxation is superior to regulation because a taxation scheme can be designed to produce any desired progressive impact \citep{Masciandaro:2013aa}. In principle, marginal tax rates can be set so that they reflect the marginal cost of reducing SR. Several authors have recently advocated for a taxation of SR \citep{Cooley:2009aa,Acharya:2010aa,Adrian:2011aa,Markose:2012aa,Acharya:2013aa,Zlatic:2014aa}, while in the real world regulation policies are being put in place due to the inherent difficulties of measuring SR \citep{BIS:2010aa}. In this context several measures for SR have recently been proposed that focus (mainly) on statistics of losses, accompanied by a potential shortfall during periods of synchronised behaviour where many institutions are simultaneously distressed \citep{Adrian:2011aa,Acharya:2010aa,Brownlees:2012aa,Huang:2012aa}. None of these measures, however, take cascading failure directly into account.

SR is predominantly a network property of liability networks \citep{Battiston:2012aa,Thurner:2013aa}. Recent econometric studies indicate that network measures could potentially serve as early warning indicators for crises \citep{Caballero:2012aa,Billio:2012aa,Minoiu:2013aa}. Different financial network topologies will have different probabilities for contagion and systemic collapse, given the link density and the financial conditions of nodes are the same \citep{Roukny:2013aa}. In this sense the management of SR becomes a technical problem of reshaping the topology of financial networks \citep{Haldane:2011aa}. The goal is to do this in a way that  neither reduces the credit provision capacity, nor the transaction volume of the financial system. Data on the topology of liability networks is available to many central banks. Several studies on historical data show typical scale-free connectivity patterns in liability networks \citep{Upper:2002aa,Boss:2004aa,Boss:2005aa,Soramaki:2007aa,Cajueiro:2009aa,Bech:2010aa,Martinez-Jaramillo:2014aa}, including overnight markets \citep{Iori:2008aa}, financial flows \citep{Kyriakopoulos:2009aa} and mutual cross holdings \citep{Huang:2013aa}. As a network property, SR can be (precisely) quantified by using network metrics \citep{Battiston:2012aa,Thurner:2013aa}. In particular, a relative network measure (DebtRank) can be assigned to all nodes in a financial network that specifies the fraction of SR that they contribute to the system (institution- or node-specific SR) \citep{Thurner:2013aa}. As shown later, it is natural to extend the notion of node-specific SR to individual liabilities between two counterparties (liability-specific SR) and to individual transactions (transaction-specific SR). 

In this paper we introduce a novel approach for the management of SR in financial networks. First, we develop a risk measure to quantify the marginal contribution of individual liabilities in financial networks to the overall SR. Second, we use this risk measure to design an incentive scheme where banks pay a Pigovian tax -- the ``Systemic Risk Tax'' (SRT) -- on each transaction, which is proportional to the increase in overall SR that it would cause. Following this approach, financial institutions would internalise their externality, as they are ``taxed'' according to their marginal contribution to overall SR. This incentive scheme leads to a self-organised reduction of SR in the following way: Market participants looking for credit will try to avoid this tax by looking for credit opportunities that do not increase SR and are thus tax-free. As a result, the network rearranges toward a topology that, in combination with the financial conditions of individual institutions, will lead to a \emph{de facto} elimination of SR. This is due to the fact that with the new topology cascading failures can no longer occur. With the help of an agent-based model (ABM), we show that financial institutions react to the SRT by rearranging the financial network over time such that overall SR is indeed drastically reduced. A number of ABMs have been used recently to study interactions between the financial system and the real economy, focusing on destabilising feedback loops between the two sectors \citep{Delli-Gatti:2009aa,Battiston:2012ab,Tedeschi:2012aa,Porter:2014aa}. We test the proposed SRT within the framework of the CRISIS macro-financial model\footnote{\url{http://www.crisis-economics.eu}}. In this ABM we run the financial system in three modes. The first reflects the situation today, where banks do not care about their systemic importance and where interbank loans are traded with an ``interbank offered rate'' that is dynamically formed in the interbank market. This interest rate only reflects the creditworthiness of the borrowing counterparty, and does not take SR into account. The second mode introduces the SRT. In this mode, the effective interest rate (interest rate + SRT) reflects both the creditworthiness of the borrowing counterparty and the SR increase associated with each transaction. For comparison, in a third mode we implement a FTT on \emph{all} transactions (Tobin-like tax) that does not depend on the SR increase associated with transactions and hence does not have any network restructuring effect.

\section{The systemic risk tax} \label{srt} 
The SRT is a levy placed on a financial transaction to offset the SR increase associated with that transaction. We show that SR associated with a transaction can be quantified by the DebtRank methodology, which was originally suggested as a recursive method to determine the systemic importance of nodes within financial networks \citep{Battiston:2012aa}. It is a quantity that measures the fraction of the total economic value (\cref{ecovalue}) in the network that is potentially affected by the default and distress of a node or a set of nodes, see \cref{debtrank_section}. For simplicity's sake let us think of the nodes in financial networks as banks. By $L_{ij}(t)$ we denote the liability (exposure\footnote{Note that the entries in $L_{ij}(t)$ are the liabilities bank $i$ has towards bank $j$. We use the convention to write liabilities in the rows (second index) of $L$. If the matrix is read column-wise (transpose of $L$) we get the assets or loans banks hold with each other.}) network of a given financial system at a given moment. $L_{ij}(t) = \sum_k l_{ijk}(t)$ is the sum of all loans $l_{ijk}(t)$ that bank $j$ currently extends to bank $i$. $C_{i}(t)$ is the capital of bank $i$ at time $t$. If bank $i$ defaults and cannot repay its loans, bank $j$ loses the loans $L_{ij}(t)$. If $j$ does not have enough capital available to cover the loss, $j$ also defaults. Given $L_{ij}(t)$ and $C_i(t)$, the DebtRank $R_i(t)=R_i(L_{ij}(t),C_i(t))$ of bank $i$ can be computed, see \cref{debtrank}.

DebtRank has the precise meaning of economic loss (in Euros) that is caused by the distress or default of a node \citep{Battiston:2012aa}. This precise meaning of the DebtRank allows us to define the ``expected systemic loss'' for the entire economy. Assuming that we have $B$ banks in the system, the expected systemic loss can be approximated by 
\begin{equation}
	EL^{\rm syst}(t) = V(t) \sum_{i=1}^{B} p_i(t) R_i(t) \quad, \label{EL}
\end{equation}
with $p_i(t)$ the probability of default of node $i$, and $V(t)$ the combined economic value of all nodes at time $t$. That this is an excellent approximation has been demonstrated in \citep{Poledna:2015aa}. For the derivation, see \cref{el_approx}. 

$R_i(t)$ measures the fraction of the total economic value (\cref{ecovalue}) that is potentially affected by node $i$. In general, $p_i(t)$ is not known and can, in principle, also depend on the particular topology of various financial networks. Since $R_i$ denotes the risk of financial contagion from the liability network $L_{ij}(t)$, the probability of default $p_i(t)$ should not explicitly depend on $L_{ij}(t)$. However, $p_i(t)$ can, in principle, depend on other networks, like the network of overlapping portfolios. Besides overlapping portfolios there are a number of reasons why default correlation exists, e.g. external events can trigger joint defaults of firms in the same geographic region or sector \citep{Hull:2012aa}. Note that we assume in \cref{EL} that $R_i$ denotes the risk of financial contagion and all other factors that lead to default correlations are comparably small (second order). Thus we calculate the total expected loss by summing the expected losses across banks. However, summing the expected losses across banks in general does not have the meaning of total expected loss because it ignores the joint probability of default. If the default correlation is known, additional terms containing the joint probability of default and the impact of a group can be added to \cref{EL}, see \cref{discussion}.

To calculate the marginal contributions to the expected systemic loss, we start by defining the \emph{net liability network} $L^{\rm net}_{ij}(t)=\max[0,L_{ij}(t)-L_{ji}(t)]$. After we add a specific liability $L_{mn}(t)$, we denote the liability network by 
\begin{equation}
	L^{(+mn)}_{ij}(t) = L^{\rm net}_{ij}(t) + \sum_{m,n}\delta_{im}\delta_{jn} L_{mn}(t) \quad, 
\end{equation}
where $\delta_{ij}$ is the Kronecker symbol. The marginal contribution of the specific liability $L_{mn}(t)$ on the expected systemic loss is 
\begin{multline}
	\Delta^{(+mn)} EL^{\rm syst}(t) = \\ 
	= \sum_{i=1}^{B} p_i(t) \left(V^{(+mn)}(t)R^{(+mn)}_i(t) - V(t)R_i(t) \right) \, , \label{marginal_effect}
\end{multline}
where $R^{(+mn)}_i(t)=R_i(L^{(+mn)}_{ij}(t),C_i(t))$ is the DebtRank of the liability network and $V^{(+mn)}(t)$ the total economic value with the added liability $L_{mn}(t)$. Clearly, a positive $\Delta^{(+mn)} EL^{\rm syst}(t)$ means that $L_{mn}(t)$ increases the total SR. 

Finally, the marginal contribution of a single loan (or a transaction leading to that loan) can be calculated. We denote a loan of bank $i$ to bank $j$ by $l_{ijk}$. The liability network changes to 
\begin{equation}
	L^{(+k)}_{ij}(t) = L^{\rm net}_{ij}(t) + \sum_{m,n,k} \delta_{im}\delta_{jn} \delta_{kk} l_{mnk}(t) \quad. \label{Lwithoutloan} 
\end{equation}
Since $i$ and $j$ can have a number of loans at a given time $t$, the index $k$ numbers a specific loan between $i$ and $j$. The marginal contribution of a single loan (transaction) $\Delta^{(+k)} EL^{\rm syst}(t)$ is obtained by substituting $L^{(+mn)}_{ij}(t)$ by $L^{(+k)}_{ij}(t)$ in \cref{marginal_effect}. In this way every existing loan in the financial system, as well as every hypothetical one, can be evaluated with respect to its marginal contribution to overall SR.

The central idea of the SRT is to tax every transaction between any two counterparties that increases SR in the system. The size of the tax is proportional to the increase of the expected systemic loss that this transaction adds to the system as seen at time $t$. The SRT for a transaction $l_{ijk}(t)$ between two banks $i$ and $j$ is given by
\begin{multline}
	SRT_{ij}^{(+k)}(t) = \zeta \max \\
	\left[0, \sum_i p_i(t) \left(V^{(+k)}(t)R^{(+k)}_i(t) - V(t)R_i(t) \right) \right] \quad. \label{srteq_simple} 
\end{multline}
Note that we assume in \cref{srteq_simple} that defaults occur only on the maturity date of the loan. For simplicity's sake we do not discount. To allow defaults at any time, valuation can be done similarly to credit risk models as for example for credit default swaps \citep{Hull:2000aa,Hull:2001aa,Hull:2012aa}\footnote{
\begin{eqnarray*}
	SRT_{ij}^{(+k)}(t) &=& \zeta \max \left[0, \int^{t+T}_t d\tau \, v(\tau) \times \right. \nonumber \\
	& \times &\left. \sum_i \hat p_i(\tau) \left(V^{(+k)}(t)R^{(+k)}_i(t) - V(t)R_i(t) \right) \right] \quad. \label{srteq} 
\end{eqnarray*}
Here $\hat p_i(\tau)$ is the default probability density of node $i$ at time $\tau$, and $v(\tau)$ the present value (at time $t$) of 1 Euro received at time $\tau$. The default probability density is defined as $\hat p_i(t) = h(t)\exp^{-\int^{\tau}_0 h(\tau)d \tau}$, where $h(t)$ is the hazard rate. The duration $T$ of the loan is from $t$ until $t+T$ and $R_i(t)$ is computed at time $t$.}. 

$\zeta$ is a proportionality constant that specifies how much of the generated expected systemic loss is taxed. $\zeta=1$ means that 100\% of the expected systemic loss will be charged. $\zeta<1$ means that only a fraction of the true SR increase is added on to the tax due from the institution responsible. $\zeta$ can be chosen such that the efficiency (total transaction volume) of the financial system is kept the same as it would be in the untaxed world. We show below that this is indeed the case.

\section{The model to test the ability of the systemic risk tax to reduce systemic risk} \label{model}
\begin{figure}
	\begin{center}
		\includegraphics[width=.99\columnwidth]{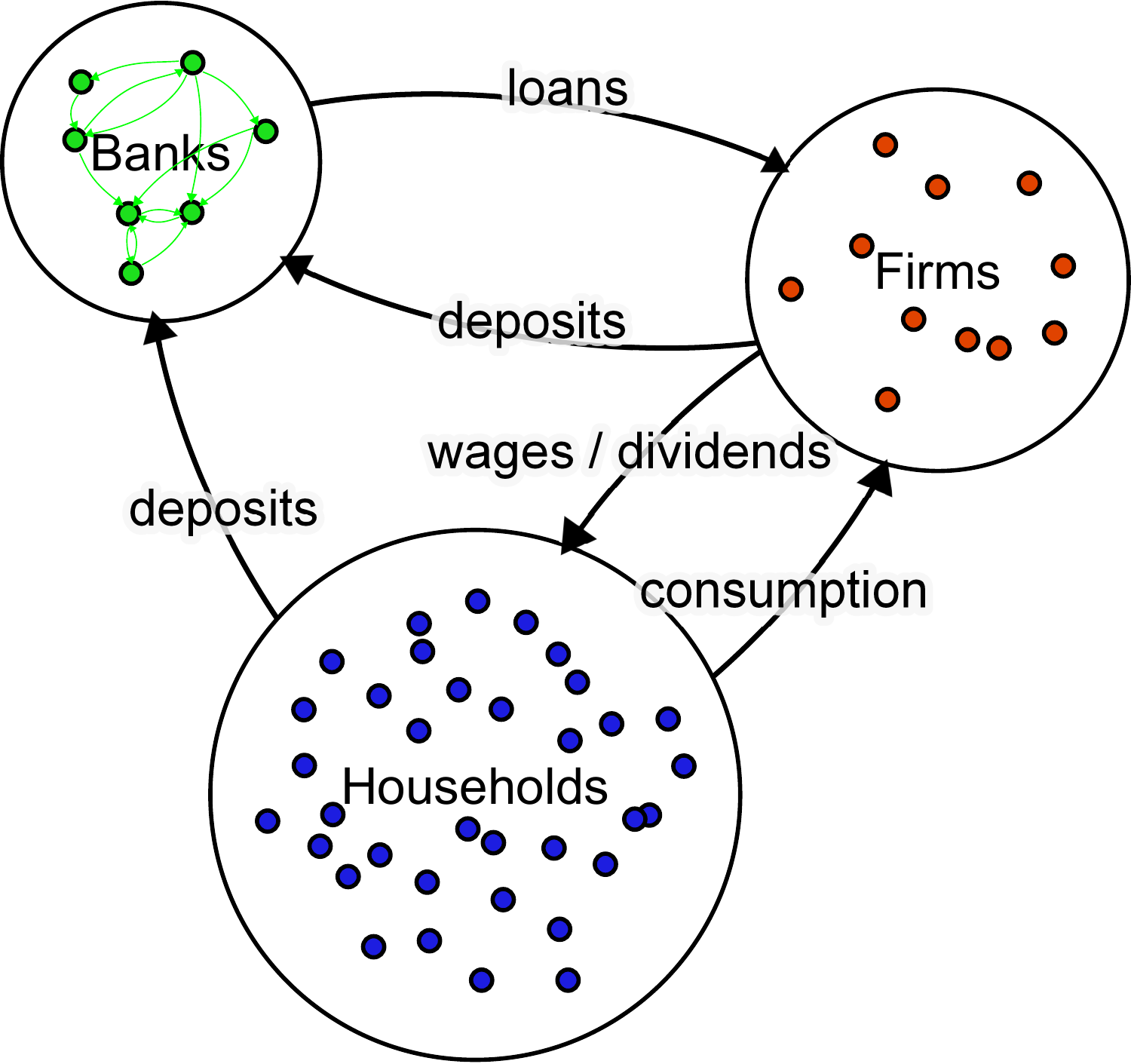} 
	\end{center}
	\caption{Schematic overview of the model structure showing the three agent types (banks, firms, and households), and their interactions. Firms pay dividends to their owners, and wages (financed through income and loans) to their workers. Households consume goods produced by the firms. Households and firms deposit money in banks, banks grant loans to the firms.} \label{model_fig} 
\end{figure}
To test the economic and financial implications of the SRT we use the CRISIS macro-financial model. This is an economic simulator that combines a well-studied macroeconomic ABM \citep{Gaffeo:2008aa,Delli-Gatti:2008aa,Delli-Gatti:2011aa} with an ABM of financial markets. We use a modified version of the macroeconomic model of \citet{Delli-Gatti:2011aa}, which additionally has an interbank market and is a \emph{closed}, stock-flow consistent economic system that allows no in- or out-flows of cash. Here we give a short description of the model, for a comprehensive description, see \citet{Delli-Gatti:2011aa} or \citet{Gualdi:2013aa} and for the modifications, see \cref{model_details}.

In the model, there are three types of agents: households, banks, and firms, as depicted in \cref{model_fig}. The agents interact on four different markets: 
\begin{enumerate}[(i)]
	\item Firms and banks interact on the credit market. 
	\item Banks interact with banks on the interbank market. 
	\item Households and firms interact on the job market.
	\item Households and firms interact on the consumption goods market. 
\end{enumerate}
Banks hold all firms' and households' cash as deposits. Households are randomly assigned as owners of firms and banks (share-holders). 
Agents repeat the following sequence of decisions at each time step:  
\begin{enumerate}[1.]
	\item firms define labour and capital demand,
	\item banks raise liquidity for loans,
	\item firms allocate capital for production (labour),
	\item households receive wages, decide on consumption and saving,
	\item firms and banks pay dividends, firms with negative liquidity go bankrupt,
	\item banks and firms repay loans,
	\item banks raise additional liquidity to manage unanticipated cash needs. 
\end{enumerate}
Households owning firms or banks receive dividends as income. All other households earn salaries for work done for firms. Banks and firms pay $20\%$ of their profits as dividends. 

\subsection{The agents} 
We give a short description of the agents; for more details on the agents and their interactions, see \citet{Delli-Gatti:2011aa}, \citet{Gualdi:2013aa}, and \cref{model_details}. 

\subsubsection{Households} There are $H$ households of which there exist two types: firm owners, and workers. Each of them has a personal account $A_{j,b}(t)$ at one of the $B$ banks. $j$ indexes the worker, $b$ the bank. Household accounts are randomly assigned to banks. Workers apply for jobs at the $F$ different firms. If hired, they receive a fixed income $w$ per time step, and supply a fixed labour productivity $\alpha$. Firm owners receive their income through dividends from their firm's profits. At each time step every household spends a fixed percentage $c$ of its current deposit account on the consumption market. Households compare prices of consumption goods from $z$ randomly chosen firms and buy the cheapest.

\subsubsection{Firms} There are $F$ firms producing perfectly substitutable consumption goods. At every time step $t$ firms compute an expected demand for the next time step $D_i(t+1)$, and an estimated price $P_i(t+1)$ (subscripts label the firm), based on a rule that takes into account both excess demand/supply and the deviation of the price $P_i(t)$ from the average price at the present time step. Each firm computes the number of required workers to supply the expected demand. If the wages for the respective workforce exceed the firm's current liquidity, it applies for a loan. Firms approach $n$ randomly chosen banks and choose the loan with the most favourable rate. If this rate exceeds a threshold rate $r^{\rm max}$, the firm only asks for $\phi$ percent of whatever loan was originally requested. Based on the outcome of this loan request, firms re-evaluate the required workforce, and hire or fire the necessary number of workers. Firms sell the goods on the consumption goods market. Firms go bankrupt if they have negative liquidity after the goods market has closed. Each of the bankrupted firm's debtors (banks) incurs a capital loss in proportion to their investment in the company. Firm owners of bankrupted firms are personally liable, and their personal account is divided by the debtors \emph{pro rata}. They immediately (next time step) start a new company. Their initial estimates for $D_i(t+1)$ and $P_i(t+1)$ equals the respective current averages in the population.

\subsubsection{Banks} 
\begin{figure}
	\begin{center}
		\includegraphics[width=.99\columnwidth]{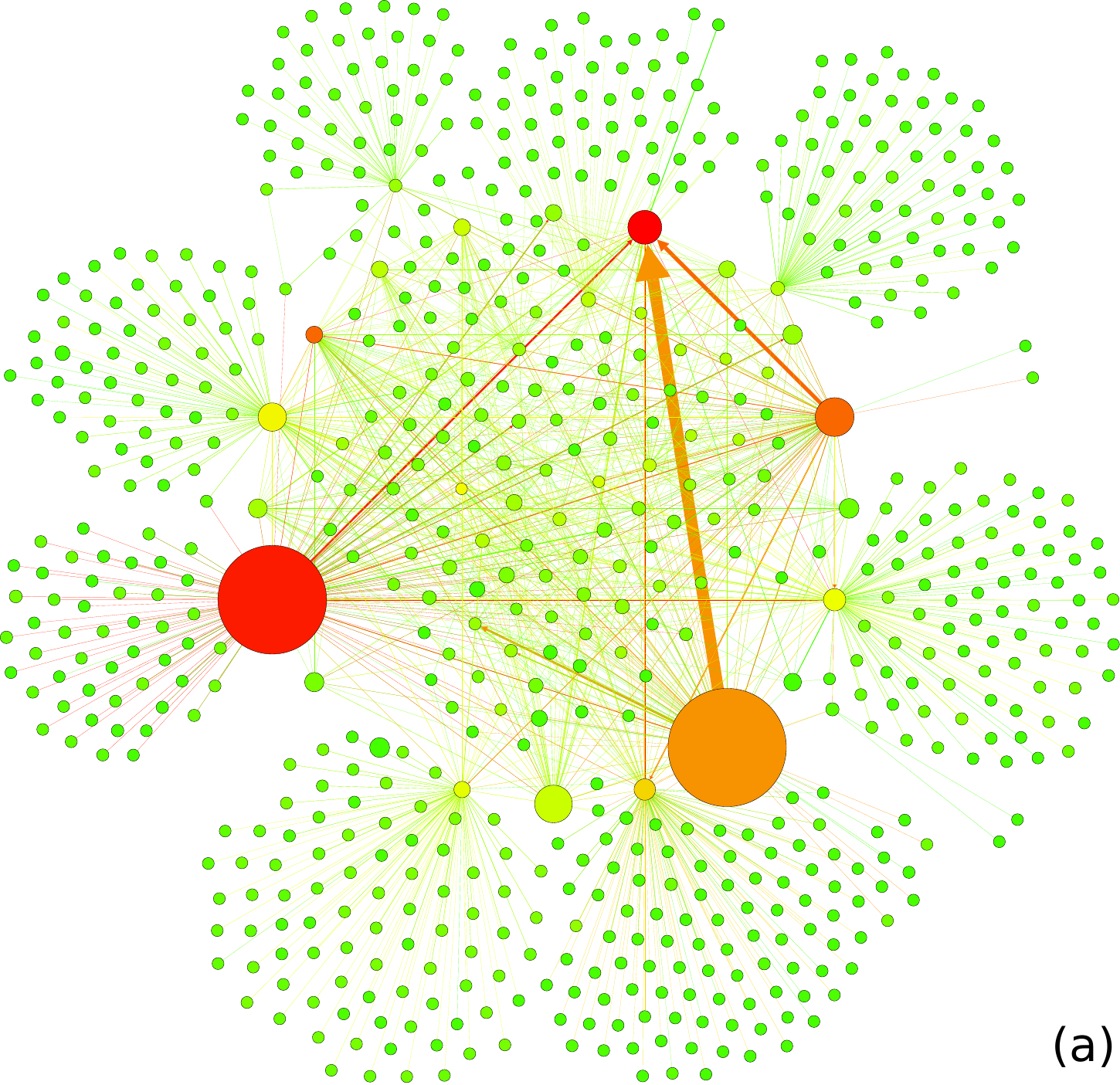} \\
		\includegraphics[width=.49\columnwidth]{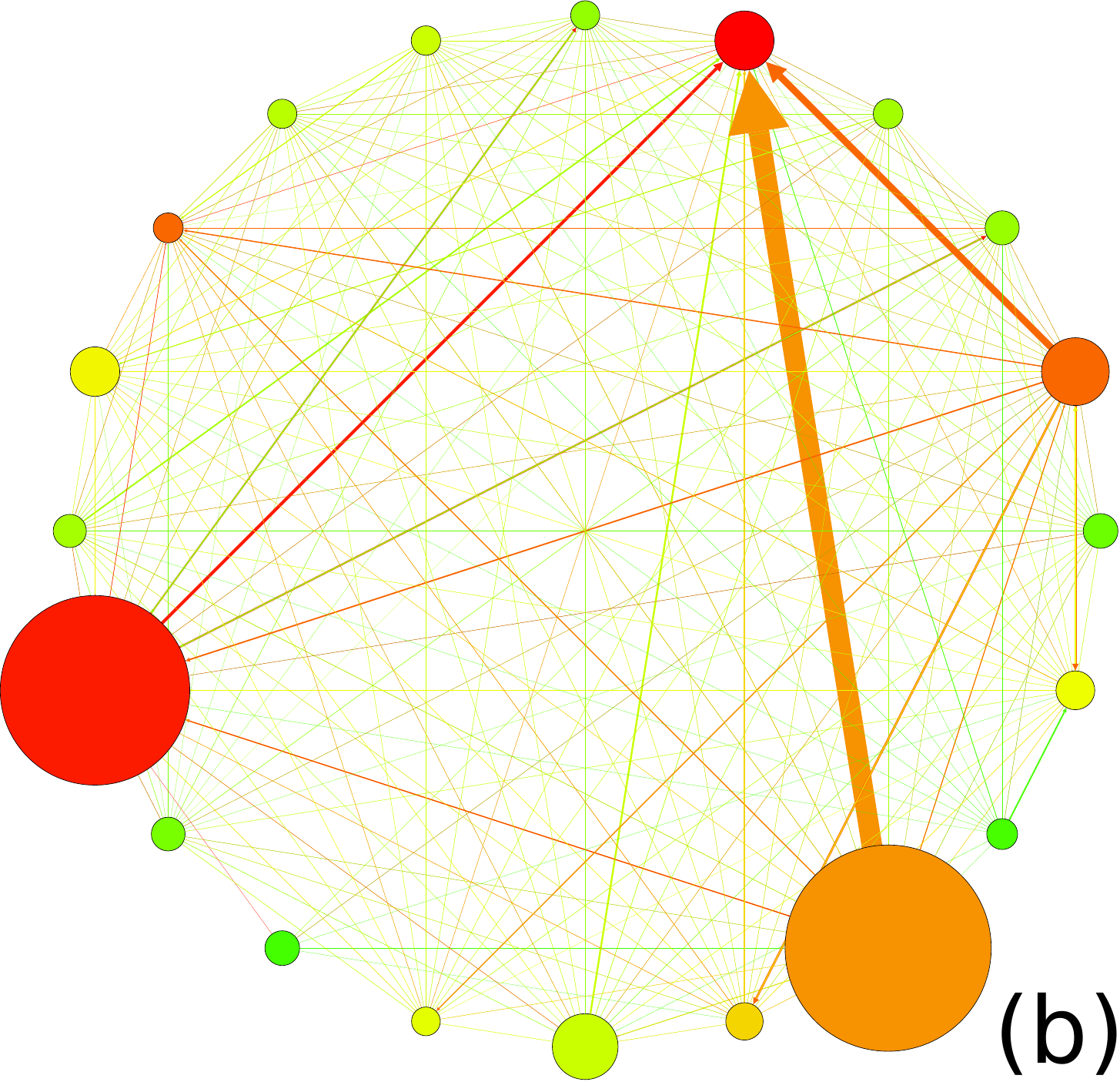} 
		\includegraphics[width=.49\columnwidth]{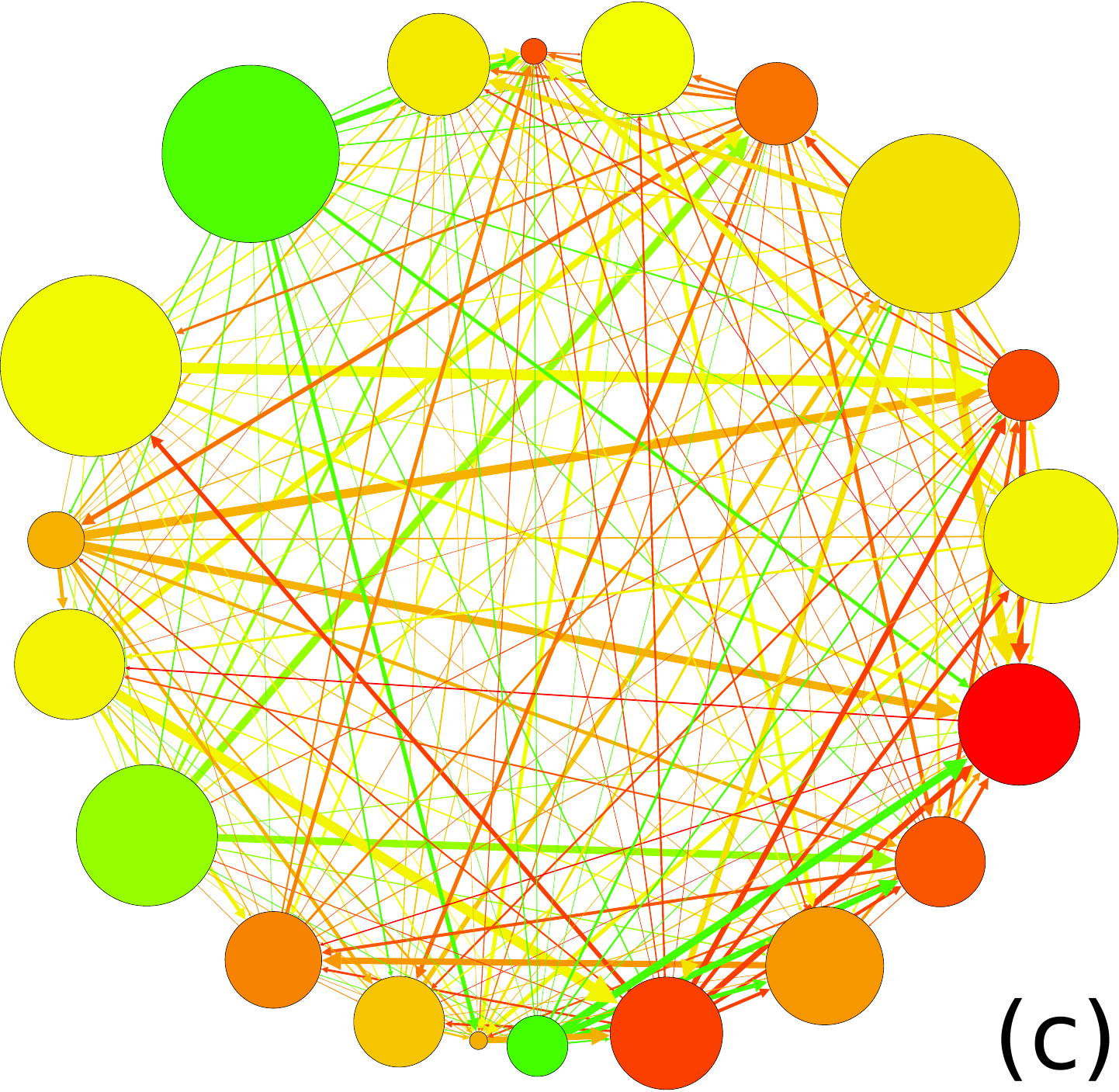} \\
		\includegraphics[width=.49\columnwidth]{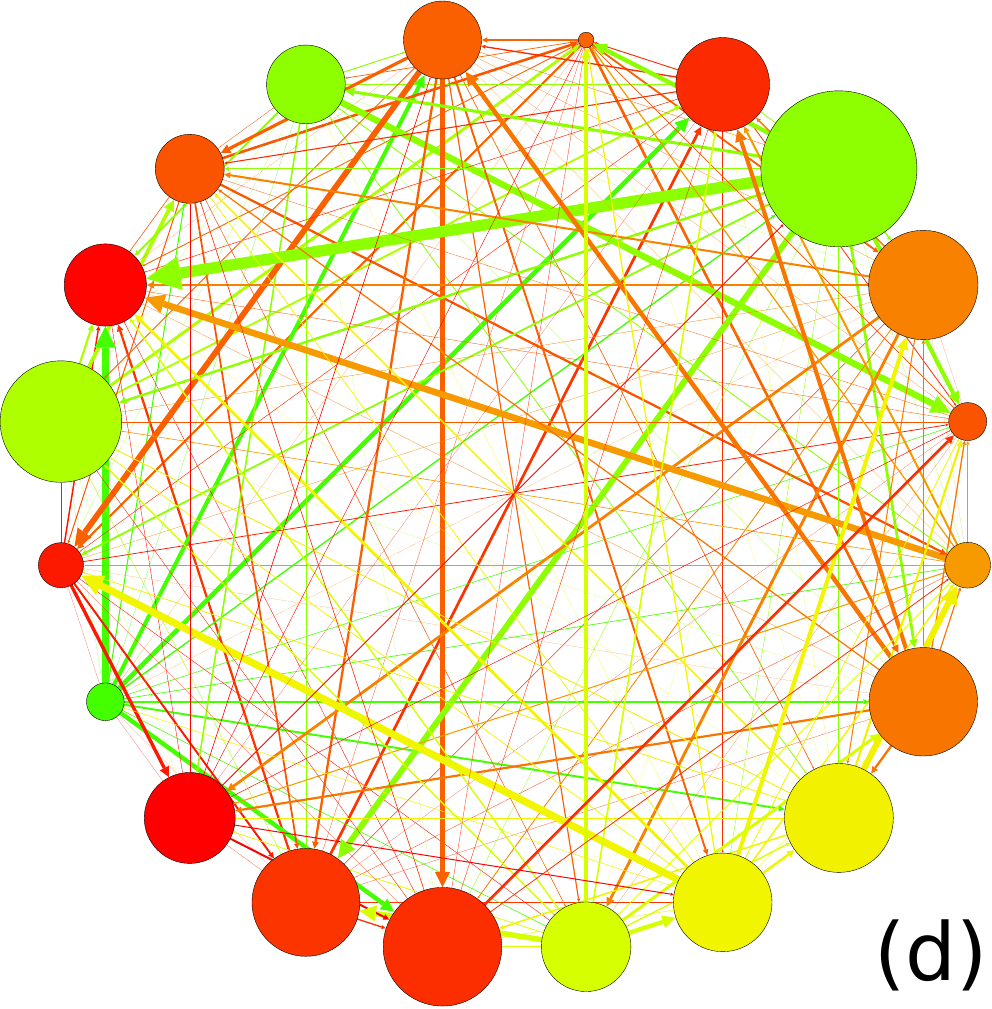} 
		\includegraphics[width=.49\columnwidth]{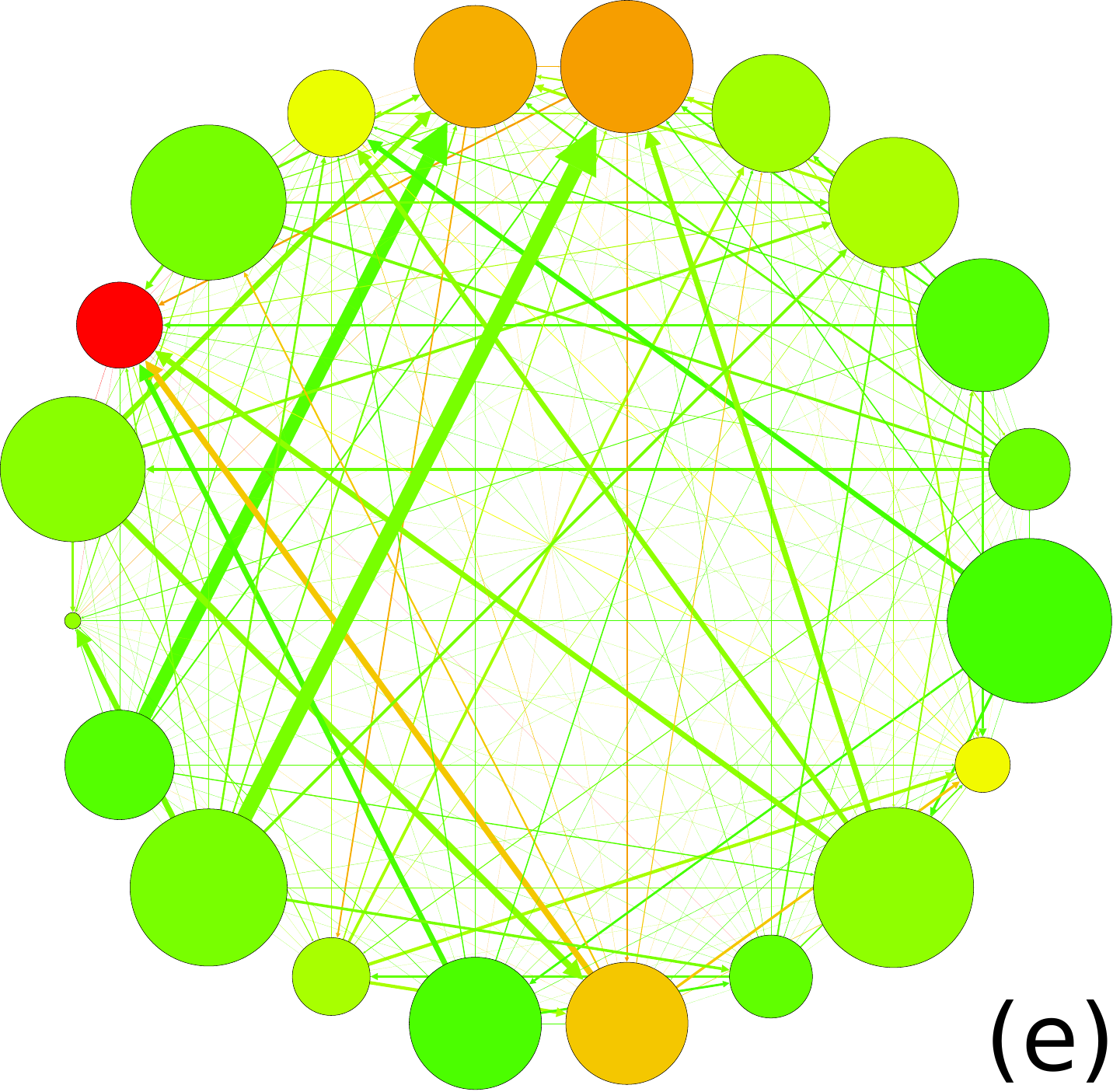} 
	\end{center}
	\caption{Banking network. (a) Austrian interbank network at the end of the first quarter of 2006, (b) the $20$ largest banks of the Austrian interbank network only, (c) banking network of the ABM without a tax, (d) with the FTT, and (e) with the SRT. Nodes (banks) are coloured according to their systemic importance $R_{i}$, from systemically important banks (red) to unimportant ones (green). The node-size represents the capitalisation of the banks. Width of the links are the liabilities of the banks in the interbank network and the colour is according to the source.} \label{network} 
\end{figure}

There are $B$ banks that offer firm loans at rates that take into account the individual specificity of banks (modelled by a uniformly distributed random variable), and the firms' creditworthiness quantified by their leverage ratio (see \cref{model_details}). Firms pay a credit risk premium according to their creditworthiness, which is modelled by a monotonically increasing function of their financial fragility. Banks try to provide requested loans and grant them if they have enough liquid resources. If they do not have enough cash, they approach other banks in the interbank market to obtain the necessary amount. If a bank does not have enough cash and cannot raise the full amount for the requested firm loan on the interbank market it does not pay out the loan. Interbank and firm loans have the same duration. Additional refinancing costs of banks remain with the firms. Each time step firms and banks repay $\tau$ percent of their outstanding debt (principal plus interest). If banks have excess liquidity they offer it on the interbank market for a nominal interest rate. The interbank market is modelled after an electronic marketplace where, in principle, all participants can enter into business relationships. In the model, banks choose the interbank offer with the most favourable rate. This does not mean that the emerging interbank network is fully connected. Emerging interbank networks are shown in \cref{network} and (weighted) degree distributions can be found in \cref{degree_dist}. Interbank rates $r_{ij}(t)$ offered by bank $i$ to bank $j$ take into account the specificity of bank $i$, and the creditworthiness (leverage ratio) of bank $j$. If a firm goes bankrupt the respective creditor bank writes off the respective outstanding loans as defaulted credits. If the bank does not have enough equity capital to cover these losses, it defaults. Following a bank default an iterative default-event unfolds for all interbank creditors. This may trigger a cascade of bank defaults. For simplicity's sake, we assume no recovery for interbank loans. This assumption is reasonable in practice for short term liquidity \citep{Cont:2013aa}. A cascade of bankruptcies happens within one time step. After the last bankruptcy is taken care of the simulation is stopped.

\subsection{Implementation of the systemic risk tax and the Tobin tax} A systemic risk premium, in form of the SRT, is imposed on all interbank transactions. Before entering a desired loan contract ${l}_{ijk}(t)$, the credit seeking banks $i$ can get quotes for the $SRT^{(+k)}_{ij}(t)$ rates from the central bank, for various offering banks $j$. They choose the interbank offer from bank $j$ with the smallest total rate, which is composed of $r^{\rm total}_{ij}(t) = r_{ij}(t)+ SRT^{(+k)}_{ij}(t)$. All other transactions are exempted from the SRT. In contrast to current market practice, the effective interest rate reflects both the creditworthiness of the borrowing counterparty and the SR increase associated with each transaction. The SRT is collected in a bailout fund. The SRT is calculated according to \cref{srteq_simple}. We approximate $p_i(t)$ by the financial fragility, defined by the borrower's leverage at time $t$. For more details, see \cref{model_details_ib}.

For comparison, we implement a Tobin-like \citep{Tobin:1978aa} FTT for interbank loans. We impose a constant tax rate of 0.2\% of the transaction (this is about 5\% of the interbank interest rates) on all interbank rates on offer. Other transactions are not taxed. The FTT makes lending less attractive for firms that borrow from banks requiring liquidity from the interbank market, as refinancing costs remain with the firms. 
\begin{figure*}
	\begin{center}
		\includegraphics[width=.99\columnwidth]{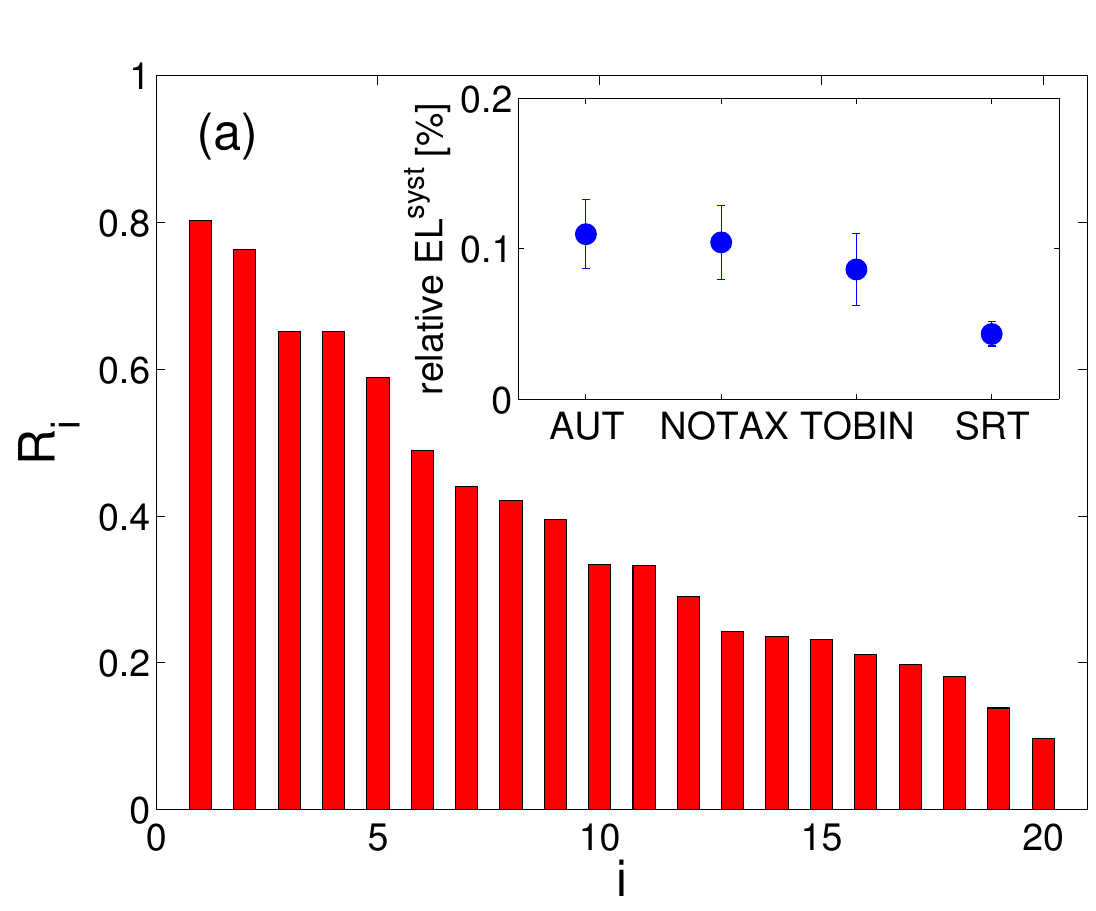} 
		\includegraphics[width=.99\columnwidth]{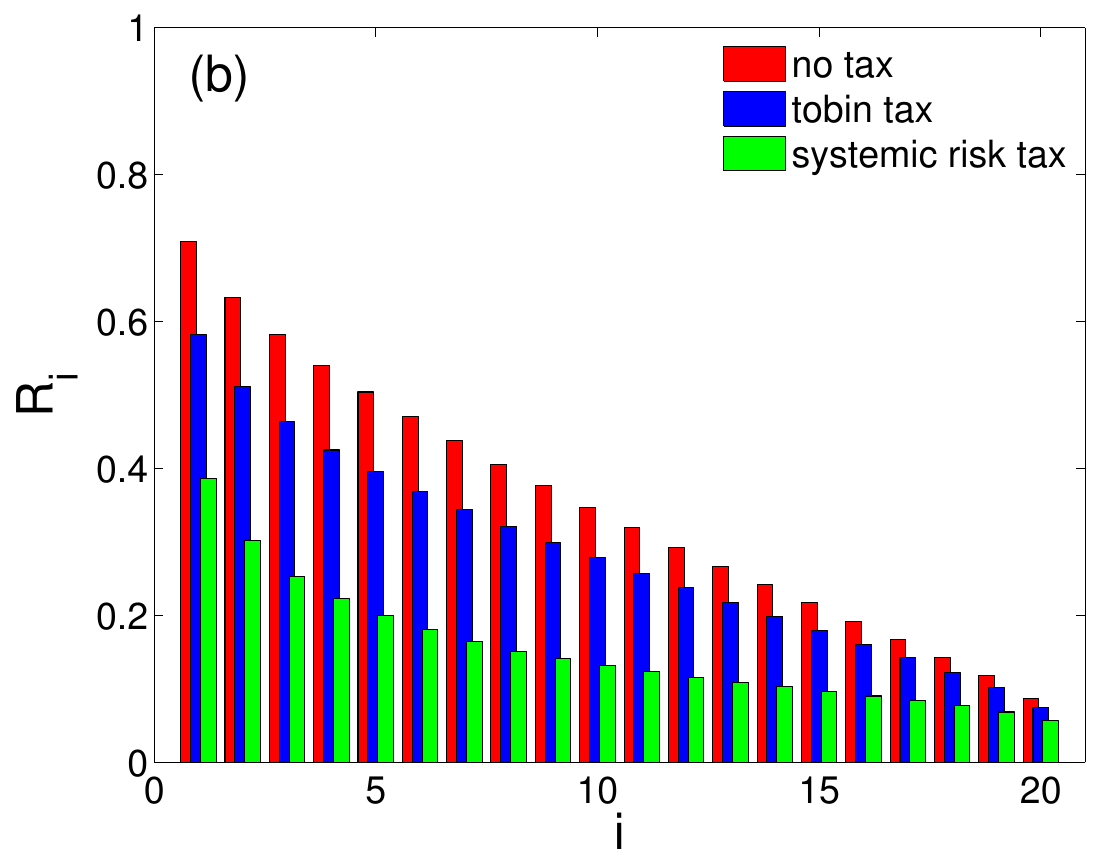} \\
		\includegraphics[width=.99\columnwidth]{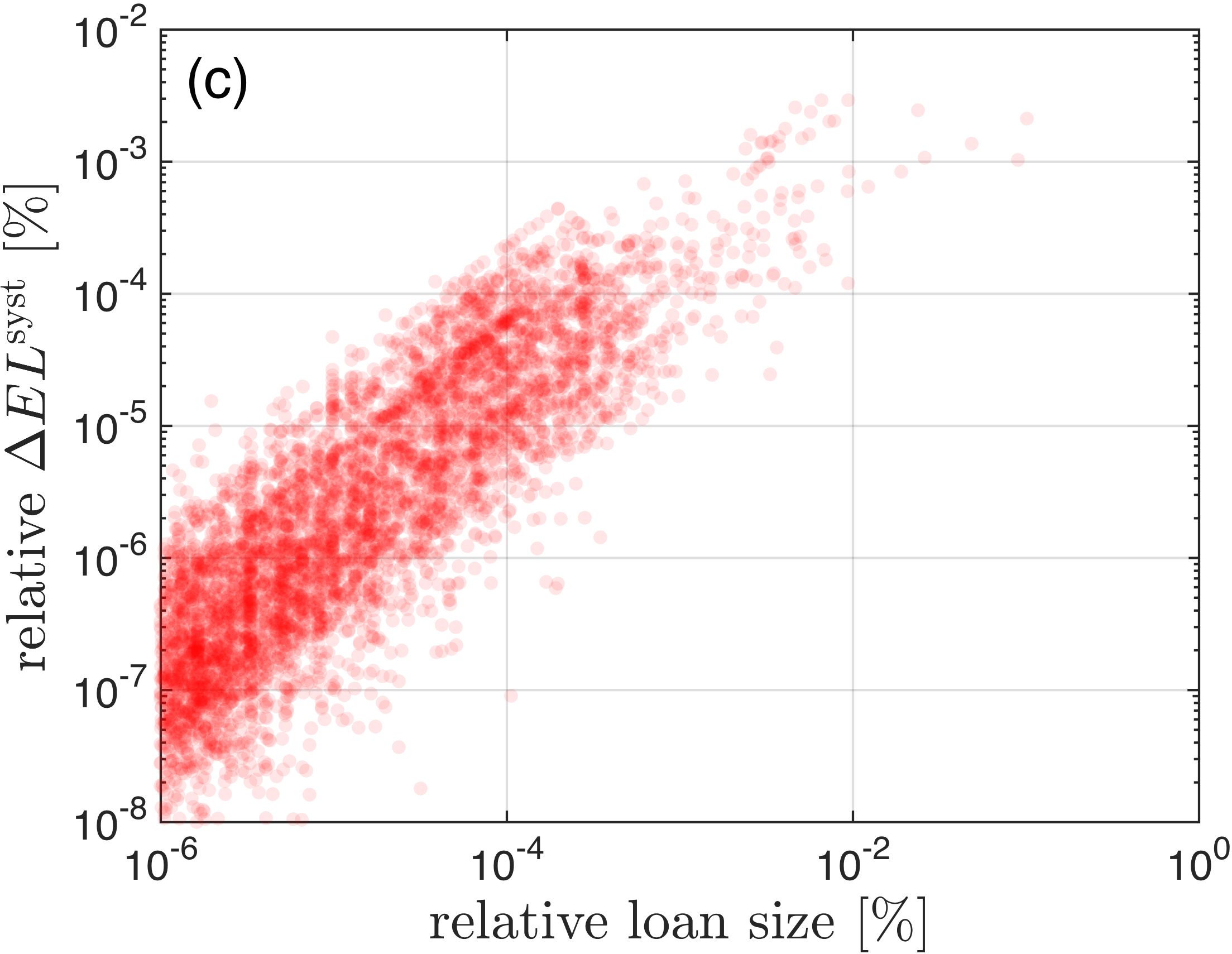} 
		\includegraphics[width=.99\columnwidth]{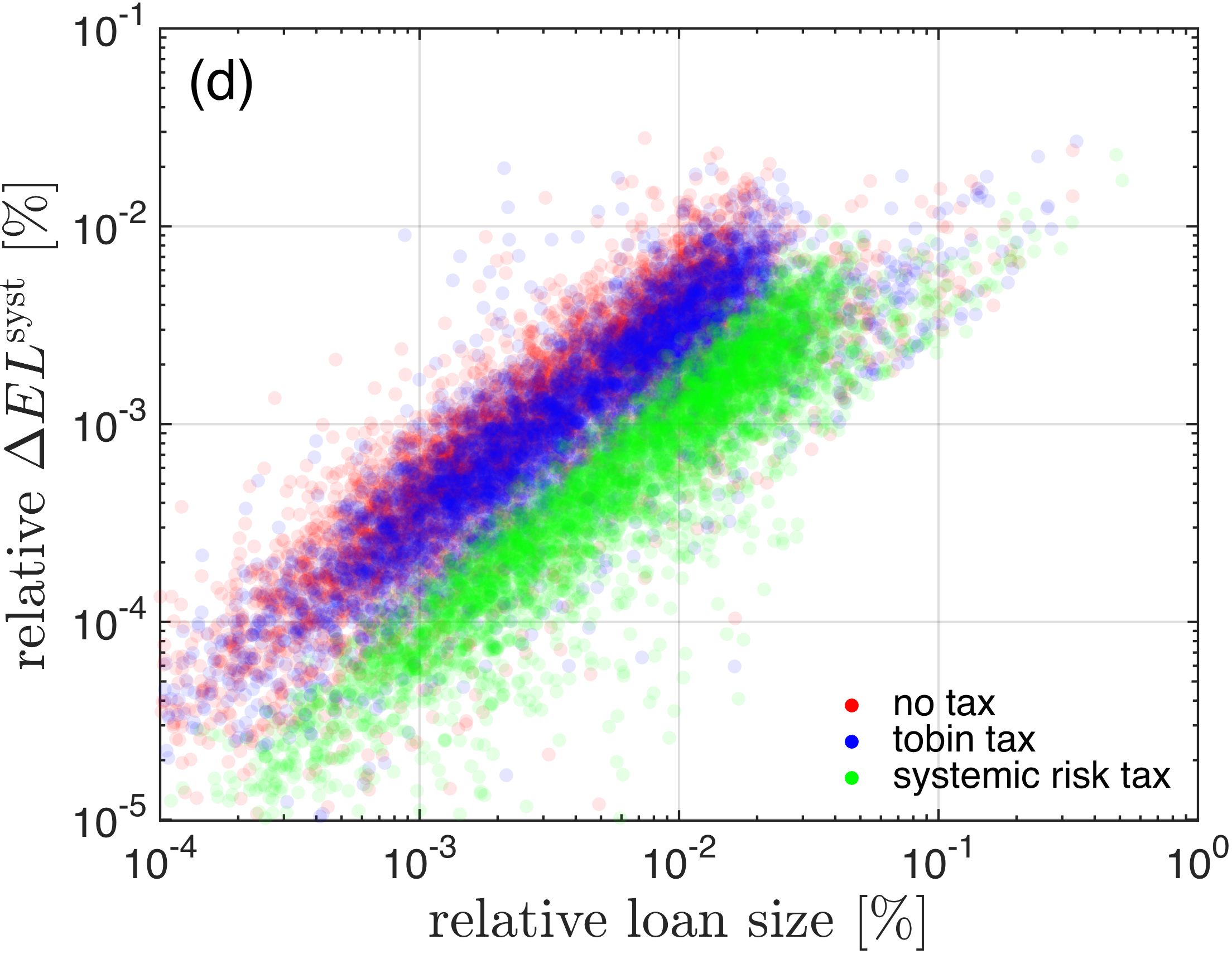} 
	\end{center}
	\caption{Expected systemic loss as measured by DebtRank, $EL^{\rm syst}_i \propto R_i$. (a) DebtRank, $R_{i}$ of the $20$ largest banks of the Austrian banking sector at the end of the first quarter of 2006. Banks are ordered by DebtRank, the most important being to the very left. Inset: Expected systemic loss from all banks for the Austrian interbank data and the three model modes. Here the SR measure is the size of a potential loss for the entire economy times the probability of that loss occurring as defined in \cref{marginal_effect}. (b) Model results for $R_{i}$: without a tax (red), with the FTT (blue), and with the SRT (green). Clearly, the SRT drastically reduces the SR contributions of individual banks. The situation without tax resembles the empirical distribution. (c) Marginal contributions on expected systemic loss $\Delta^{(+mn)} EL^{\rm syst}$ of individual interbank liabilities $L_{mn}$ vs. the relative size of interbank loans in double logarithmic scale. Every data point represents an interbank liability $L^{\rm data}_{mn}$, see \cref{data}. The loan size captures the credit risk for lenders, whereas $\Delta^{(+mn)} EL^{\rm syst}$ is the SR of the liability. (d) Marginal contributions for the simulations in the three modes. The SRT reduces SR but leaves contract sizes unchanged.} \label{data_vs_model} 
\end{figure*}

\section{Results} \label{results} 
We implement the above model in MATLAB for $B=20$ banks, $F=100$ firms, and $H=1300$ households. The model is run in three modes, without any tax, with the SRT, and with a Tobin-like FTT. Results are averages over $10,000$ independent, identical simulations across $500$ time steps. We set $\zeta=0.02$ (see \cref{srt}), and for the Tobin-like FTT we impose a constant tax rate of 0.2\% of the transaction. For different tax rates for the Tobin-like FTT and an alternative mode in which the SRT is set to the true increase in SR associated with each transaction ($\zeta=1$), see \cref{tobin}. Additionally in \cref{tobin} there is a short discussion of the effect of the SRT on the network properties.

We compare model results to historical, anonymised, and linearly transformed interbank liability data provided by the Austrian Central Bank (OeNB), see \cref{data}. In \cref{network}(a) we show a snapshot of the Austrian interbank network at the end of the first quarter of 2006. Nodes symbolise the banks of the Austrian banking system and links represent their lending relations (weighted by the liabilities). Nodes are coloured according to their systemic importance $R_{i}$, from systemically important banks (red) to unimportant ones (green). The node-size represents the capitalisation of banks and the width of the links symbolises the liabilities. In \cref{network}(b) we show the $20$ largest banks of the Austrian interbank network. Clearly, the $20$ largest banks contribute most to overall SR (red and orange dots).
\Cref{network}(c) shows results from the ABM without a tax, (d) with the FTT, and (e) with the SRT. The SRT effectively reduces the spreading of SR by preventing systemically important nodes from lending. This can be seen from the fact that there are only green links in \cref{network}(e). In the snapshot of the Austrian interbank network and in the model without the SRT numerous red links are clearly visible. In \cref{data_vs_model}(a) we show SR as measured by DebtRank $R_{i}$. In particular, we show $R_{i}$ for the $20$ largest banks (according to total assets) of the Austrian banking sector at the end of the first quarter of 2006. Here we calculate $R_{i}$ from $L^{\rm data}_{ij}(t)$ (see \cref{data}), in \cref{data_vs_model}(b) we use the net liability network $L^{\rm net}_{ij}(t)$. Banks are ordered by their DebtRank, the systemically most important one is to the very left, the least important one to the very right. The ABM results for $R_{i}(t)$ are presented in \cref{data_vs_model}(b): without a tax (red), with the FTT (blue) and with the SRT (green). The shown distributions are averages over $10,000$ independent simulations. Clearly, the SRT drastically reduces the SR contributions of individual banks. The situation without tax resembles the empirical distribution remarkably well. In \cref{data_vs_model}(c) the marginal contributions on expected systemic loss from \cref{marginal_effect} are presented for all individual interbank liabilities $L^{\rm data}_{mn}$, as a function of the relative size of interbank loans. Every data point represents a single interbank liability $L^{\rm data}_{mn}$ from bank $m$ to $n$. Interbank loans are themselves power-law distributed (not shown), which is known empirically \citep{Boss:2005aa}. The loan size captures the credit risk for lenders, whereas $\Delta^{(+mn)} EL^{\rm syst}$ is the SR contribution of the liability. \Cref{data_vs_model}(d) shows the corresponding marginal contributions for the ABM simulations for the three modes. It is clearly visible that the SRT reduces the SR contribution of liabilities by approximately an order of magnitude (note the log scale), but leaves contract-sizes practically unchanged. 
\begin{figure}
	\begin{center}
		\includegraphics[width=.99\columnwidth]{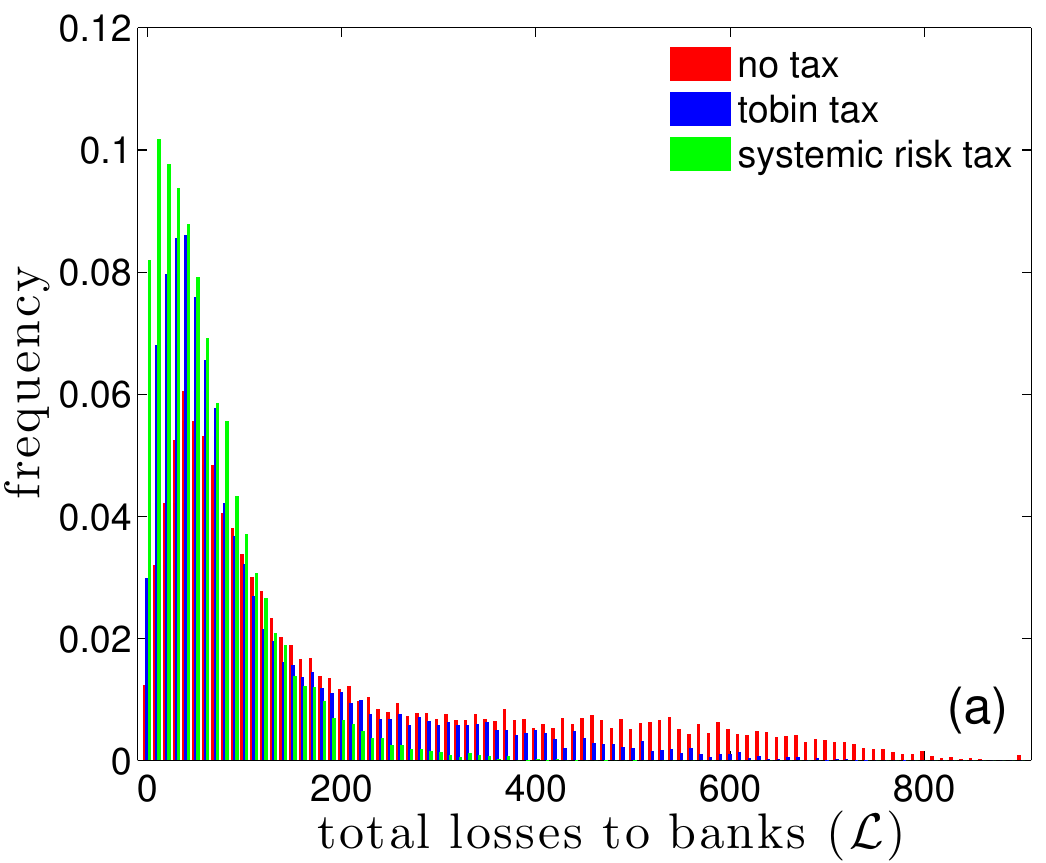} 
		\includegraphics[width=.99\columnwidth]{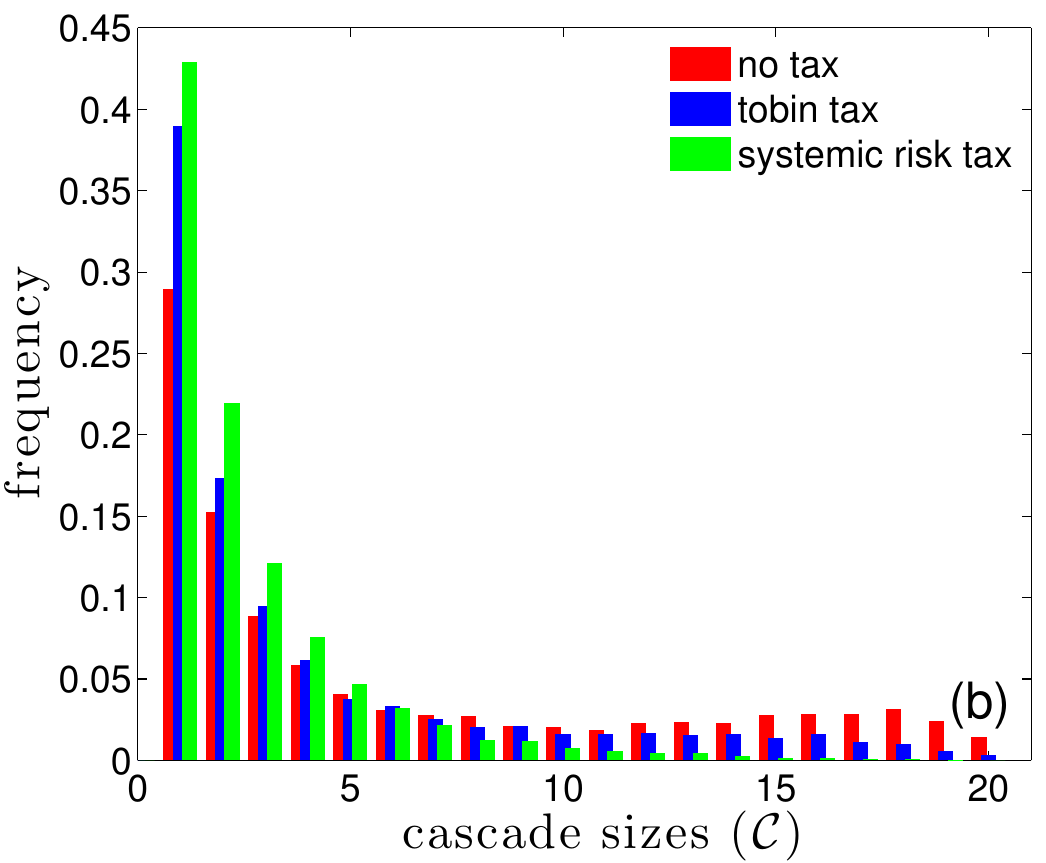} 
		\includegraphics[width=.99\columnwidth]{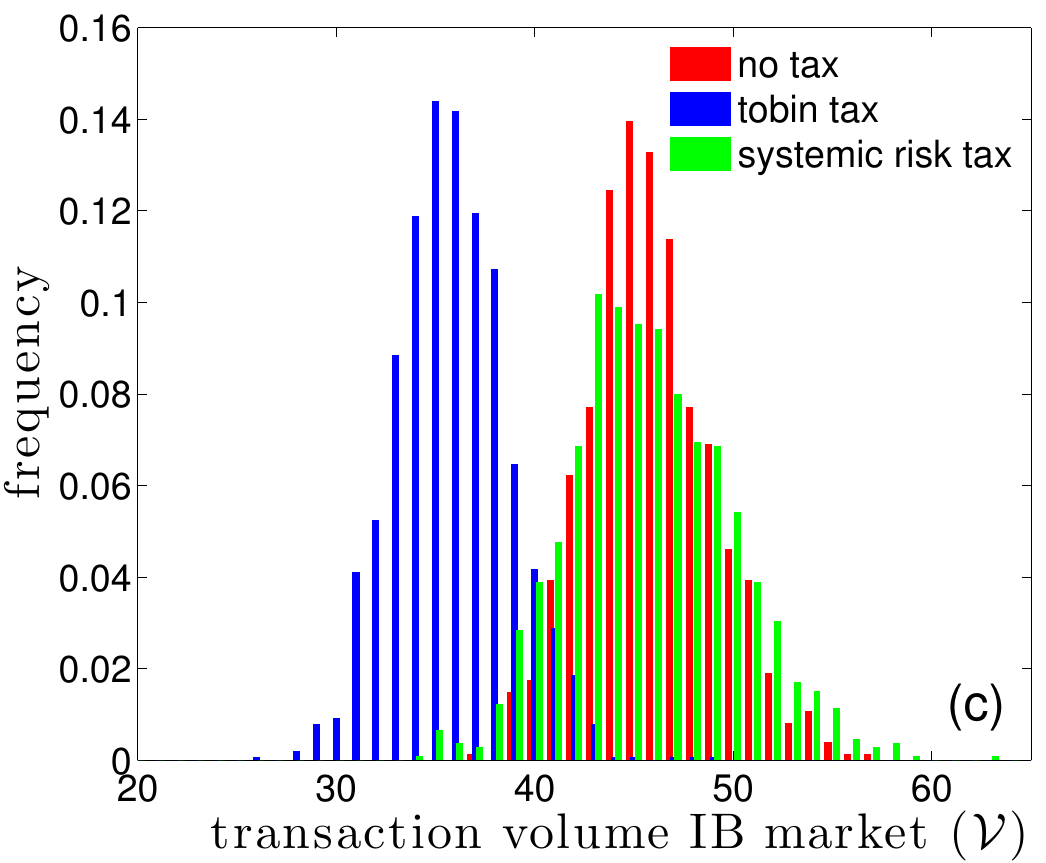} 
	\end{center}
	\caption{Comparison of no financial transaction tax (red) on interbank loans, with systemic risk tax (green), and Tobin tax (blue). (a) Distribution of total losses to banks ${\cal L}$, (b) distribution of cascade sizes ${\cal C}$ of defaulting banks, and (c) distribution of total transaction volume in the interbank market ${\cal V}$. $10,000$ independent, identical simulations, each with $500$ time steps, $20$ banks.} \label{hist_losses} 
\end{figure}
The effects of the SRT and the FTT on total losses to banks ${\cal L}$ (see \cref{risk_measures}) that occur as a consequence of bank defaults are shown in \cref{hist_losses}(a). Clearly, the mode without tax (red) produces fat tails in the loss distributions of the banking sector. The Tobin tax slightly reduces losses. The SRT gets rid of big losses in the system completely (green). The remaining losses reflect those from firm defaults, which represents the economic risk in the system. Note that economic risk can hardly be managed. This elimination of losses on the interbank market is due to the fact that under the SRT the possibility for cascading defaults is drastically reduced. This is seen in \cref{hist_losses}(b), where the distributions of cascade sizes $\mathcal{S}$ (see \cref{risk_measures}) for the three modes are compared. While the untaxed mode produces considerable cascade sizes of up to $20$ banks, the maximum cascade sizes under the SRT is less than half. The Tobin tax more or less follows the untaxed case. As mentioned above, the interbank loan sizes are practically unchanged under the SRT. This is also true for the total transaction volume $\mathcal{V}$ (see \cref{risk_measures}) in the interbank market, as can be seen in \cref{hist_losses}(c), where the distributions of transaction volumes at time step 100 are shown. Obviously, the situation for the SRT (green) is very similar to the untaxed case (red), whereas the transaction volume is drastically reduced in the FTT scenario (blue), as expected. 

\section{Discussion} \label{discussion}
We extend the notion of SR to individual liabilities within a financial network and show with empirical data of nation-wide interbank liabilities that this is indeed feasible. The notion of liability-specific SR allows us to quantify the marginal contribution of every financial transaction to overall SR. We propose a Pigovian tax (SRT) on every SR-increasing transaction, proportional to the marginal contribution to overall SR. By trying to avoid the SRT, financial institutions effectively rearrange the financial network over time, such that cascading failures can no longer occur. This process leads to a sustainable, self-organised and self-stabilising reduction of SR.

The notion of liability-specific SR is based on the probability of default and the impact of a failure of a financial institution, as measured by DebtRank. A central idea of this paper is to separate \emph{default risk} from \emph{contagion risk}. Contagion risk is the risk that a default by one institution leads to defaults of other institutions. DebtRank denotes the risk of financial contagion from the interbank liability network. The default risk of a financial institution depends on its financial condition and on the economic situation in general. In principle, it can also depend on financial networks, for example, the network of overlapping portfolios or the network of firms and regions or industries. However, if network contributions to the probability of default are small (second order), it becomes possible to separate default risk from contagion risk, as given by DebtRank, in a meaningful way. Otherwise, it is necessary to replace or generalise DebtRank with a methodology that includes the network of overlapping portfolios and other relevant networks. 

In \citet{Poledna:2015aa} we show that DebtRank and the ideas presented here can be generalised for multi-layer networks to quantify and reduce SR originating from different financial markets, such as derivatives, foreign exchange and securities. In future work we will conduct an empirical study on the network of overlapping portfolios and its implications for SR. Joint defaults can also be taken into account for example by copula methods. Once the default correlation has been estimated, it is possible to include the joint probability of default straightforwardly in the present framework. Joint defaults can be included by considering the joint probability of default of a group of financial institutions and by calculating the DebtRank for this group. Thus, additional terms containing the joint probability of default and the impact of a group can be added to \cref{EL,srteq_simple}.

We test the SRT within the framework of the CRISIS macro-financial model. The model produces SR-profiles of banks that are practically identical to those of actual interbank liability data. Even on the level of individual transactions the model is fully compatible with the empirical data (see also \citep{Poledna:2015aa}). The SRT drastically reduces the probability of a financial collapse due to restructured liability networks that minimise the size of cascading failure. The tax is implemented in a simple way: an agent would like to make a transaction (with a given counterparty) and expresses this interest by announcing it (and the envisioned counterparty) to the central bank. The latter computes the SR increase associated with the transaction based on the knowledge of the present state of the entire liability network and the capitalisation of its agents. The SR-increase is then presented to the agent as a tax (SRT) for that particular transaction. If the SR-increase is zero, then it is tax-free. The agent can now look for other counterparties to carry out exactly the same transaction. The agent will therefore typically screen several possible counterparties and then decide on the one with the lowest tax. Once the agent decides to carry out the transaction, it is executed and the tax is paid to the central bank or the government.

We show explicitly that SR is, to a large extent, a network property. We show that the SRT is able to restructure financial liability networks without loss of transaction volume in the financial market. For an explicit comparison we implement and test a Tobin-like tax that taxes all transactions regardless of their SR contributions. The Tobin-like tax does not restructure networks and only reduces SR because it also drastically reduces transaction volume in the system. This is damaging as it makes the system less efficient; the loss of efficiency materialises as expensive credit for the real economy. We tested an alternative mode in which the SRT is set to the true increase in SR associated with the transaction, and not only a fraction ($\zeta=1$). This alternative leads to much more homogeneous SR-spreading across all agents, and makes the system even safer, see \cref{data_vs_model}(b) and \cref{hist_srt}(d), however, it is done at the cost of much reduced transaction volume.

An obvious alternative to the SRT are bank taxes for SIFIs as recently suggested by several authors \citep{Cooley:2009aa,Acharya:2010aa,Adrian:2011aa,Markose:2012aa,Acharya:2013aa,Zlatic:2014aa},  or alternatively, to increase capital requirements for them by imposing SIFI surcharges as proposed in Basel III \citep{BIS:2010aa}. An immediate problem of a bank tax or capital surcharges, compared to the SRT, is that financial institutions sometimes have no control over their systemic importance. For instance, in case of publicly traded securities, such as bonds, financial institutions have no authority to decide who holds them and thus no influence on their systemic importance. In recent work the effects of the Basel III regulation framework on SR has been studied in the same ABM environment \citep{Poledna:2016aa}. Results indicate that capital surcharges for SIFIs can reduce SR, but must be larger than specified in Basel III to have a measurable impact, and thus cause a loss of efficiency.

We close with a remark on the policy relevance of the SRT. We think that the concept that network-related SR can be managed most efficiently by restructuring the underlying network topologies, is generally true. We further believe that the concept of the SRT as presented here is directly policy relevant. In particular, the incentive scheme introduced to internalise the externalities that lead to SR, can be directly transferred to banking regulation. Technically, the requirements for its implementation would include an electronic market run by a central bank or another central authority. This electronic market would work in the same way as an airline reservations system, i.e. by holding a quote for a limited amount of time. The computational requirements for central banks to compute the SRT for several thousand banks on a minute by minute basis is by today’s standards a technical triviality. Central banks would have to record transactions in real-time; this is done as a matter of course for several financial markets routinely, for instance at the Banco de M\'{e}xico \citep{Solorzano-Margain:2013aa,Poledna:2015aa}. There are no privacy issues; obtaining information about the portfolios of other banks from the SRT quotes is impossible, in the same way as it is impossible to infer the entries of a real valued matrix from the eigenvector associated to the largest real eigenvalue. The problem that generally applies to any FTT -- that it should be implemented globally to avoid free riding -- also applies in the context of the SRT. 

We finally stress that the current market practice of not pricing SR into transaction-costs effectively amounts to an implicit subsidy for those with the highest contribution to overall SR, and an effective tax on those with the lowest. In that sense, the SRT can be seen as an insurance premium that keeps the public neutral with respect to the SR that is introduced by the non-optimal liability networks. 

\section*{Acknowledgements}
We thank P. Klimek for many stimulating conversations. We acknowledge financial support from EC FP7 projects CRISIS, agreement no. 288501 (65\%), LASAGNE, agreement no. 318132 (15\%) and MULTIPLEX, agreement no. 317532 (20\%).

\bibliographystyle{unsrtnat}
\bibliography{econophysics}

\begin{thebibliography}{75}
\providecommand{\natexlab}[1]{#1}
\providecommand{\url}[1]{\texttt{#1}}
\expandafter\ifx\csname urlstyle\endcsname\relax
  \providecommand{\doi}[1]{doi: #1}\else
  \providecommand{\doi}{doi: \begingroup \urlstyle{rm}\Url}\fi

\bibitem[{The Economist}(2013)]{ecobailout}
{The Economist}.
\newblock What {A}ngela isn't saying, August 2013.

\bibitem[{The Economist}(2007{\natexlab{a}})]{ecoresession}
{The Economist}.
\newblock America's vulnerable economy, November 2007{\natexlab{a}}.

\bibitem[{The Economist}(2007{\natexlab{b}})]{ecocrunch}
{The Economist}.
\newblock {CSI}: credit crunch, October 2007{\natexlab{b}}.

\bibitem[Klimek et~al.(2015)Klimek, Poledna, Farmer, and
  Thurner]{Klimek:2014aa}
P.~Klimek, S.~Poledna, J.D. Farmer, and S.~Thurner.
\newblock To bail-out or to bail-in? answers from an agent-based model.
\newblock \emph{Journal of Economic Dynamics and Control}, 50:\penalty0
  144--154, 2015.

\bibitem[Tobin(1978)]{Tobin:1978aa}
James Tobin.
\newblock A proposal for international monetary reform.
\newblock Technical Report 506, Cowles Foundation for Research in Economics,
  Yale University, 1978.

\bibitem[Summers and Summers(1989)]{Summers:1989aa}
L.~H. Summers and V.~P. Summers.
\newblock When financial markets work too well: A cautious case for a
  securities transactions tax.
\newblock Technical Report~12, Columbia - Center for Futures Markets, 1989.

\bibitem[McCulloch and Pacillo(2011)]{McCulloch:2011aa}
Neil McCulloch and Grazia Pacillo.
\newblock The tobin tax a review of the evidence.
\newblock Technical Report 1611, Department of Economics, University of Sussex,
  Jan 2011.

\bibitem[Matheson(2012)]{Matheson:2012aa}
Thornton Matheson.
\newblock Security transaction taxes: issues and evidence.
\newblock \emph{International Tax and Public Finance}, 19\penalty0
  (6):\penalty0 884--912, 2012.

\bibitem[Aikman et~al.(2013)Aikman, Haldane, and Kapadia]{Aikman:2013aa}
D.~Aikman, A.~G. Haldane, and S.~Kapadia.
\newblock Operationalising a macroprudential regime: Goals, tools and open
  issues.
\newblock \emph{Financial Stability Journal of the Bank of Spain}, 24, 2013.

\bibitem[{Bank of England}(2011)]{Bank-of-England:2011aa}
{Bank of England}.
\newblock Instruments of macroprudential policy.
\newblock Technical report, Bank of England, 2011.

\bibitem[{Bank of England}(2013)]{Bank-of-England:2013aa}
{Bank of England}.
\newblock The financial policy committee's powers to supplement capital
  requirements: a draft policy statement.
\newblock Technical report, Bank of England, 2013.

\bibitem[{Bank for International Settlements}(2010)]{BIS:2010aa}
{Bank for International Settlements}.
\newblock \emph{{Basel III}: A global regulatory framework for more resilient
  banks and banking systems}.
\newblock Bank for International Settlements, 2010.

\bibitem[Georg(2011)]{Georg:2011aa}
Co-Pierre Georg.
\newblock {Basel III} and systemic risk regulation - what way forward?
\newblock Technical Report~17, Working Papers on Global Financial Markets,
  2011.

\bibitem[Duffie and Singleton(2012)]{Duffie:2012aa}
D.~Duffie and K.J. Singleton.
\newblock \emph{Credit Risk: Pricing, Measurement, and Management}.
\newblock Princeton Series in Finance. Princeton University Press, 2012.
\newblock ISBN 9781400829170.

\bibitem[{Bank for International Settlements}(1988)]{BIS:1988aa}
{Bank for International Settlements}.
\newblock \emph{International convergence of capital measurement and capital
  standards}.
\newblock Bank for International Settlements, Basel, 1988.

\bibitem[{Bank for International Settlements}(2006)]{BIS:2006aa}
{Bank for International Settlements}.
\newblock \emph{International Convergence of Capital Measurement and Capital
  Standards: A Revised Framework Comprehensive Version}.
\newblock Bank for International Settlements, Basel, 2006.

\bibitem[Balin(2008)]{Balin:2008aa}
Bryan~J. Balin.
\newblock {Basel I, Basel II}, and emerging markets: A nontechnical analysis.
\newblock \emph{Available at SSRN 1477712}, 2008.

\bibitem[De~Bandt and Hartmann(2000)]{De-Bandt:2000aa}
Olivier De~Bandt and Philipp Hartmann.
\newblock Systemic risk: A survey.
\newblock Technical report, CEPR Discussion Papers, 2000.

\bibitem[Eisenberg and Noe(2001)]{Eisenberg:2001aa}
Larry Eisenberg and Thomas~H. Noe.
\newblock Systemic risk in financial systems.
\newblock \emph{Management Science}, 47\penalty0 (2):\penalty0 236--249, 2001.

\bibitem[Minsky(1992)]{Minsky:1992aa}
Hyman~P. Minsky.
\newblock The financial instability hypothesis.
\newblock \emph{The Jerome Levy Economics Institute Working Paper}, 74, 1992.

\bibitem[Fostel and Geanakoplos(2008)]{Fostel:2008aa}
Ana Fostel and John Geanakoplos.
\newblock Leverage cycles and the anxious economy.
\newblock \emph{American Economic Review}, 98\penalty0 (4):\penalty0 1211--44,
  2008.

\bibitem[Geanakoplos(2010)]{Geanakoplos:2010aa}
John Geanakoplos.
\newblock The leverage cycle.
\newblock In D.~Acemoglu, K.~Rogoff, and M.~Woodford, editors, \emph{NBER
  Macro-economics Annual 2009}, volume~24, page 165. University of Chicago
  Press, 2010.

\bibitem[Adrian and Shin(2008)]{Adrian:2008aa}
Tobias Adrian and Hyun~S. Shin.
\newblock Liquidity and leverage.
\newblock Tech. Rep. 328, Federal Reserve Bank of New York, 2008.

\bibitem[Brunnermeier and Pedersen(2009)]{Brunnermeier:2009aa}
Markus Brunnermeier and Lasse Pedersen.
\newblock Market liquidity and funding liquidity.
\newblock \emph{Review of Financial Studies}, 22\penalty0 (6):\penalty0
  2201--2238, 2009.

\bibitem[Thurner et~al.(2012)Thurner, Farmer, and Geanakoplos]{Thurner:2012aa}
S.~Thurner, J.D. Farmer, and J.~Geanakoplos.
\newblock Leverage causes fat tails and clustered volatility.
\newblock \emph{Quantitative Finance}, 12\penalty0 (5):\penalty0 695--707,
  2012.

\bibitem[Caccioli et~al.(2012)Caccioli, Bouchaud, and Farmer]{Caccioli:2012aa}
Fabio Caccioli, Jean-Philippe Bouchaud, and J.~Doyne Farmer.
\newblock Impact-adjusted valuation and the criticality of leverage.
\newblock \emph{Risk}, 25\penalty0 (12), 2012.

\bibitem[Poledna et~al.(2014)Poledna, Thurner, Farmer, and
  Geanakoplos]{Poledna:2014ab}
Sebastian Poledna, Stefan Thurner, J.~Doyne Farmer, and John Geanakoplos.
\newblock Leverage-induced systemic risk under {Basle II} and other credit risk
  policies.
\newblock \emph{Journal of Banking \& Finance}, 42:\penalty0 199--212, 2014.

\bibitem[Aymanns and Farmer(2015)]{Aymanns:2014aa}
Christoph Aymanns and Doyne Farmer.
\newblock The dynamics of the leverage cycle.
\newblock \emph{Journal of Economic Dynamics and Control}, 50:\penalty0
  155--179, 2015.

\bibitem[Caccioli et~al.(2015)Caccioli, Farmer, Foti, and
  Rockmore]{Caccioli:2015aa}
Fabio Caccioli, J~Doyne Farmer, Nick Foti, and Daniel Rockmore.
\newblock Overlapping portfolios, contagion, and financial stability.
\newblock \emph{Journal of Economic Dynamics and Control}, 51:\penalty0 50--63,
  2015.

\bibitem[Acharya et~al.(2009)Acharya, Pedersen, Philippon, and
  Richardson]{Acharya:2009aa}
Viral Acharya, Lasse Pedersen, Thomas Philippon, and Matthew Richardson.
\newblock Regulating systemic risk.
\newblock \emph{Restoring financial stability: How to repair a failed system},
  pages 283--304, 2009.

\bibitem[Davies and Tracey(2014)]{Davies:2014aa}
Richard Davies and Belinda Tracey.
\newblock Too big to be efficient? the impact of implicit subsidies on
  estimates of scale economies for banks.
\newblock \emph{Journal of Money, Credit and Banking}, 46\penalty0
  (s1):\penalty0 219--253, 2014.

\bibitem[Acharya et~al.(2012)Acharya, Pedersen, Philippon, and
  Richardson]{Acharya:2010aa}
Viral Acharya, Lasse Pedersen, Thomas Philippon, and Matthew Richardson.
\newblock Measuring systemic risk.
\newblock Technical report, CEPR Discussion Papers, 2012.
\newblock Available at SSRN: http://ssrn.com/abstract=1573171.

\bibitem[Acemoglu et~al.(2013)Acemoglu, Ozdaglar, and
  Tahbaz-Salehi]{Acemoglu:2013aa}
Daron Acemoglu, Asuman Ozdaglar, and Alireza Tahbaz-Salehi.
\newblock Systemic risk and stability in financial networks.
\newblock Technical report, National Bureau of Economic Research, 2013.

\bibitem[Masciandaro and Passarelli(2013)]{Masciandaro:2013aa}
Donato Masciandaro and Francesco Passarelli.
\newblock Financial systemic risk: Taxation or regulation?
\newblock \emph{Journal of Banking \& Finance}, 37\penalty0 (2):\penalty0
  587--596, 2013.

\bibitem[Cooley et~al.(2009)Cooley, Philippon, Acharya, Pedersen, and
  Richardson]{Cooley:2009aa}
Thomas Cooley, Thomas Philippon, Viral Acharya, Lasse Pedersen, and Matthew
  Richardson.
\newblock Regulating systemic risk.
\newblock In Viral Acharya and Matthew~P Richardson, editors, \emph{Restoring
  Financial Stability: How to Repair a Failed System}, pages 277--303. John
  Wiley \& Sons, 2009.

\bibitem[Adrian and Brunnermeier(2011)]{Adrian:2011aa}
Tobias Adrian and Markus Brunnermeier.
\newblock Covar.
\newblock Technical report, National Bureau of Economic Research, 2011.

\bibitem[Markose et~al.(2012)Markose, Giansante, and Shaghaghi]{Markose:2012aa}
Sheri Markose, Simone Giansante, and Ali~Rais Shaghaghi.
\newblock {`Too interconnected to fail'} financial network of {US CDS} market:
  Topological fragility and systemic risk.
\newblock \emph{Journal of Economic Behavior \& Organization}, 83\penalty0
  (3):\penalty0 627--646, 2012.

\bibitem[Acharya et~al.(2013)Acharya, Pedersen, Philippon, and
  Richardson]{Acharya:2013aa}
Viral Acharya, Lasse Pedersen, Thomas Philippon, and Matthew Richardson.
\newblock Taxing systemic risk.
\newblock In J.P. Fouque and J.A. Langsam, editors, \emph{Handbook on Systemic
  Risk}, pages 226--246. Cambridge University Press, 2013.
\newblock ISBN 9781107023437.

\bibitem[Zlati{\'c} et~al.(2014)Zlati{\'c}, Gabbi, and Abraham]{Zlatic:2014aa}
Vinko Zlati{\'c}, Giampaolo Gabbi, and Hrvoje Abraham.
\newblock Reduction of systemic risk by means of pigouvian taxation.
\newblock \emph{arXiv preprint arXiv:1406.5817}, 2014.

\bibitem[Brownlees and Engle(2012)]{Brownlees:2012aa}
Christian~T Brownlees and Robert~F Engle.
\newblock Volatility, correlation and tails for systemic risk measurement.
\newblock \emph{Available at SSRN 1611229}, 2012.

\bibitem[Huang et~al.(2012)Huang, Zhou, and Zhu]{Huang:2012aa}
Xin Huang, Hao Zhou, and Haibin Zhu.
\newblock Systemic risk contributions.
\newblock \emph{Journal of financial services research}, 42\penalty0
  (1-2):\penalty0 55--83, 2012.

\bibitem[Battiston et~al.(2012{\natexlab{a}})Battiston, Puliga, Kaushik, Tasca,
  and Caldarelli]{Battiston:2012aa}
Stefano Battiston, Michelangelo Puliga, Rahul Kaushik, Paolo Tasca, and Guido
  Caldarelli.
\newblock Debtrank: Too central to fail? financial networks, the {FED} and
  systemic risk.
\newblock \emph{Scientific reports}, 2\penalty0 (541), 2012{\natexlab{a}}.

\bibitem[Thurner and Poledna(2013)]{Thurner:2013aa}
Stefan Thurner and Sebastian Poledna.
\newblock Debtrank-transparency: Controlling systemic risk in financial
  networks.
\newblock \emph{Scientific reports}, 3\penalty0 (1888), 2013.

\bibitem[Caballero(2012)]{Caballero:2012aa}
J.~Caballero.
\newblock Banking crises and financial integration.
\newblock \emph{IDB Working Paper Series No. IDB-WP-364}, 2012.

\bibitem[Billio et~al.(2012)Billio, Getmansky, Lo, and Pelizzon]{Billio:2012aa}
Monica Billio, Mila Getmansky, Andrew~W Lo, and Loriana Pelizzon.
\newblock Econometric measures of connectedness and systemic risk in the
  finance and insurance sectors.
\newblock \emph{Journal of Financial Economics}, 104\penalty0 (3):\penalty0
  535--559, 2012.

\bibitem[Minoiu et~al.(2013)Minoiu, Kang, Subrahmanian, and
  Berea]{Minoiu:2013aa}
Camelia Minoiu, Chanhyun Kang, V.~S. Subrahmanian, and Anamaria Berea.
\newblock Does financial connectedness predict crises?
\newblock Technical Report 13/267, International Monetary Fund, Dec 2013.

\bibitem[Roukny et~al.(2013)Roukny, Bersini, Pirotte, Caldarelli, and
  Battiston]{Roukny:2013aa}
Tarik Roukny, Hugues Bersini, Hugues Pirotte, Guido Caldarelli, and Stefano
  Battiston.
\newblock Default cascades in complex networks: Topology and systemic risk.
\newblock \emph{Scientific Reports}, 3\penalty0 (2759), 2013.

\bibitem[Haldane and May(2011)]{Haldane:2011aa}
Andrew~G. Haldane and Robert~M. May.
\newblock Systemic risk in banking ecosystems.
\newblock \emph{Nature}, 469\penalty0 (7330):\penalty0 351--355, 2011.

\bibitem[Upper and Worms(2002)]{Upper:2002aa}
Christian Upper and Andreas Worms.
\newblock Estimating bilateral exposures in the german interbank market: Is
  there a danger of contagion?
\newblock Technical Report~9, Deutsche Bundesbank, Research Centre, 2002.

\bibitem[Boss et~al.(2004)Boss, Summer, and Thurner]{Boss:2004aa}
M.~Boss, M.~Summer, and S.~Thurner.
\newblock Contagion flow through banking networks.
\newblock \emph{Lecture Notes in Computer Science}, 3038:\penalty0 1070--1077,
  2004.

\bibitem[Boss et~al.(2005)Boss, Elsinger, Summer, and Thurner]{Boss:2005aa}
M.~Boss, H.~Elsinger, M.~Summer, and S.~Thurner.
\newblock The network topology of the interbank market.
\newblock \emph{Quantitative Finance}, 4:\penalty0 677--684, 2005.

\bibitem[Soram{\"a}ki et~al.(2007)Soram{\"a}ki, Bech, Arnold, Glass, and
  Beyeler]{Soramaki:2007aa}
Kimmo Soram{\"a}ki, Morten~L. Bech, Jeffrey Arnold, Robert~J. Glass, and
  Walter~E. Beyeler.
\newblock The topology of interbank payment flows.
\newblock \emph{Physica A: Statistical Mechanics and its Applications},
  379\penalty0 (1):\penalty0 317--333, 2007.

\bibitem[Cajueiro et~al.(2009)Cajueiro, Tabak, and Andrade]{Cajueiro:2009aa}
Daniel~O Cajueiro, Benjamin~M Tabak, and Roberto~FS Andrade.
\newblock Fluctuations in interbank network dynamics.
\newblock \emph{Physical Review E}, 79\penalty0 (3), 2009.

\bibitem[Bech and Atalay(2010)]{Bech:2010aa}
Morten~L. Bech and Enghin Atalay.
\newblock The topology of the federal funds market.
\newblock \emph{Physica A: Statistical Mechanics and its Applications},
  389\penalty0 (22):\penalty0 5223--5246, 2010.

\bibitem[Mart\'{i}nez-Jaramillo et~al.(2014)Mart\'{i}nez-Jaramillo,
  Alexandrova-Kabadjova, Bravo-Benitez, and
  Sol{\'o}rzano-Margain]{Martinez-Jaramillo:2014aa}
Seraf\'{i}n Mart\'{i}nez-Jaramillo, Biliana Alexandrova-Kabadjova, Bernardo
  Bravo-Benitez, and Juan~Pablo Sol{\'o}rzano-Margain.
\newblock An empirical study of the mexican banking system's network and its
  implications for systemic risk.
\newblock \emph{Journal of Economic Dynamics and Control}, 40:\penalty0
  242--265, 2014.
\newblock ISSN 0165-1889.

\bibitem[Iori et~al.(2008)Iori, De~Masi, Precup, Gabbi, and
  Caldarelli]{Iori:2008aa}
Giulia Iori, Giulia De~Masi, Ovidiu~Vasile Precup, Giampaolo Gabbi, and Guido
  Caldarelli.
\newblock A network analysis of the italian overnight money market.
\newblock \emph{Journal of Economic Dynamics and Control}, 32\penalty0
  (1):\penalty0 259--278, 2008.

\bibitem[Kyriakopoulos et~al.(2009)Kyriakopoulos, Thurner, Puhr, and
  Schmitz]{Kyriakopoulos:2009aa}
F.~Kyriakopoulos, S.~Thurner, C.~Puhr, and S.~W. Schmitz.
\newblock Network and eigenvalue analysis of financial transaction networks.
\newblock \emph{The European Physical Journal B - Condensed Matter and Complex
  Systems}, 71\penalty0 (4):\penalty0 523--531, October 2009.

\bibitem[Huang et~al.(2013)Huang, Vodenska, Havlin, and Stanley]{Huang:2013aa}
Xuqing Huang, Irena Vodenska, Shlomo Havlin, and H.~Eugene Stanley.
\newblock Cascading failures in bi-partite graphs: Model for systemic risk
  propagation.
\newblock \emph{Scientific reports}, 3\penalty0 (1219), 2013.

\bibitem[{Delli Gatti} et~al.(2009){Delli Gatti}, Gallegati, Greenwald, Russo,
  and Stiglitz]{Delli-Gatti:2009aa}
Domenico {Delli Gatti}, Mauro Gallegati, Bruce~C Greenwald, Alberto Russo, and
  Joseph~E Stiglitz.
\newblock Business fluctuations and bankruptcy avalanches in an evolving
  network economy.
\newblock \emph{Journal of Economic Interaction and Coordination}, 4\penalty0
  (2):\penalty0 195--212, 2009.

\bibitem[Battiston et~al.(2012{\natexlab{b}})Battiston, Delli~Gatti, Gallegati,
  Greenwald, and Stiglitz]{Battiston:2012ab}
Stefano Battiston, Domenico Delli~Gatti, Mauro Gallegati, Bruce Greenwald, and
  Joseph~E Stiglitz.
\newblock Liaisons dangereuses: Increasing connectivity, risk sharing, and
  systemic risk.
\newblock \emph{Journal of Economic Dynamics and Control}, 36\penalty0
  (8):\penalty0 1121--1141, 2012{\natexlab{b}}.

\bibitem[Tedeschi et~al.(2012)Tedeschi, Mazloumian, Gallegati, and
  Helbing]{Tedeschi:2012aa}
Gabriele Tedeschi, Amin Mazloumian, Mauro Gallegati, and Dirk Helbing.
\newblock Bankruptcy cascades in interbank markets.
\newblock \emph{PloS one}, 7\penalty0 (12):\penalty0 e52749, 2012.

\bibitem[Gabbi et~al.(2015)Gabbi, Iori, Jafarey, and Porter]{Porter:2014aa}
Giampaolo Gabbi, Giulia Iori, Saqib Jafarey, and James Porter.
\newblock Financial regulations and bank credit to the real economy.
\newblock \emph{Journal of Economic Dynamics and Control}, 50:\penalty0
  117--143, 2015.

\bibitem[Poledna et~al.(2015)Poledna, Molina-Borboa, Mart{\'\i}nez-Jaramillo,
  van~der Leij, and Thurner]{Poledna:2015aa}
Sebastian Poledna, Jos{\'e}~Luis Molina-Borboa, Seraf{\'\i}n
  Mart{\'\i}nez-Jaramillo, Marco van~der Leij, and Stefan Thurner.
\newblock The multi-layer network nature of systemic risk and its implications
  for the costs of financial crises.
\newblock \emph{Journal of Financial Stability}, 20:\penalty0 70--81, 2015.

\bibitem[Hull(2012)]{Hull:2012aa}
John~C Hull.
\newblock \emph{Options, Futures, and Other Derivatives}.
\newblock Prentice Hall PTR, 2012.
\newblock ISBN 9780132164948.

\bibitem[Hull and White(2000)]{Hull:2000aa}
John~C Hull and Alan~D White.
\newblock Valuing credit default swaps {I}: No counterparty default risk.
\newblock \emph{The Journal of Derivatives}, 8\penalty0 (1):\penalty0 29--40,
  2000.

\bibitem[Hull and White(2001)]{Hull:2001aa}
John~C Hull and Alan~D White.
\newblock Valuing credit default swaps {II}: Modeling default correlations.
\newblock \emph{The Journal of Derivatives}, 8\penalty0 (3):\penalty0 12--21,
  2001.

\bibitem[Gaffeo et~al.(2008)Gaffeo, Delli~Gatti, Desiderio, and
  Gallegati]{Gaffeo:2008aa}
Edoardo Gaffeo, Domenico Delli~Gatti, Saul Desiderio, and Mauro Gallegati.
\newblock Adaptive microfoundations for emergent macroeconomics.
\newblock \emph{Eastern Economic Journal}, 34\penalty0 (4):\penalty0 441--463,
  2008.

\bibitem[{Delli Gatti} et~al.(2008){Delli Gatti}, Gaffeo, Gallegati, Giulioni,
  and Palestrini]{Delli-Gatti:2008aa}
D.~{Delli Gatti}, E.~Gaffeo, M.~Gallegati, G.~Giulioni, and A.~Palestrini.
\newblock \emph{Emergent Macroeconomics: An Agent-Based Approach to Business
  Fluctuations}.
\newblock New economic windows. Springer, 2008.
\newblock ISBN 9788847007253.

\bibitem[{Delli Gatti} et~al.(2011){Delli Gatti}, Desiderio, Gaffeo, Cirillo,
  and Gallegati]{Delli-Gatti:2011aa}
Domenico {Delli Gatti}, Saul Desiderio, Edoardo Gaffeo, Pasquale Cirillo, and
  Mauro Gallegati.
\newblock \emph{Macroeconomics from the Bottom-up}.
\newblock Springer Milan, 2011.

\bibitem[Gualdi et~al.(2015)Gualdi, Tarzia, Zamponi, and
  Bouchaud]{Gualdi:2013aa}
Stanislao Gualdi, Marco Tarzia, Francesco Zamponi, and Jean-Philippe Bouchaud.
\newblock Tipping points in macroeconomic agent-based models.
\newblock \emph{Journal of Economic Dynamics and Control}, 50:\penalty0 29--61,
  2015.

\bibitem[Cont et~al.(2013)Cont, Moussa, and Santos]{Cont:2013aa}
Rama Cont, Amal Moussa, and Edson Santos.
\newblock Network structure and systemic risk in banking systems.
\newblock In Jean-Pierre Fouque and Joseph~A Langsam, editors, \emph{Handbook
  of Systemic Risk}, pages 327--368. Cambridge University Press, 2013.

\bibitem[Poledna et~al.(2016)Poledna, Bochmann, and Thurner]{Poledna:2016aa}
Sebastian Poledna, Olaf Bochmann, and Stefan Thurner.
\newblock {Basel III} capital surcharges for {G-SIBs} fail to control systemic
  risk and can cause pro-cyclical side effects.
\newblock \emph{arXiv preprint arXiv:1602.03505}, 2016.

\bibitem[Solorzano-Margain et~al.(2013)Solorzano-Margain, Martinez-Jaramillo,
  and Lopez-Gallo]{Solorzano-Margain:2013aa}
Juan~Pablo Solorzano-Margain, Serafin Martinez-Jaramillo, and Fabrizio
  Lopez-Gallo.
\newblock Financial contagion: Extending the exposures network of the mexican
  financial system.
\newblock \emph{Computational Management Science}, 10\penalty0 (2-3):\penalty0
  125--155, 2013.

\bibitem[Barrat et~al.(2004)Barrat, Barth{\'e}lemy, Pastor-Satorras, and
  Vespignani]{Barrat:2004aa}
A.~Barrat, M.~Barth{\'e}lemy, R.~Pastor-Satorras, and A.~Vespignani.
\newblock The architecture of complex weighted networks.
\newblock \emph{Proceedings of the National Academy of Sciences of the United
  States of America}, 101\penalty0 (11):\penalty0 3747--3752, 03 2004.

\bibitem[{Standard \& Poor's}(2009)]{standardandpoors}
{Standard \& Poor's}.
\newblock Understanding {Standard \& Poor's} rating definitions, June 2009.

\end{thebibliography}
 
\newpage
\clearpage

\appendix

\section{Details of the model} \label{model_details} In this section we describe the extensions and modifications of the macroeconomic model of \citet{Delli-Gatti:2011aa}. The modifications include the implementation of an interbank market and a {\em closed}, stock-flow consistent economic system that allows no in- or out-flow of cash. The closed economic system is also discussed in \citet{Gualdi:2013aa}.

\subsection{The credit market} There are $B$ banks that offer firm loans at rates that take the individual specificity of banks (modelled by a uniformly distributed random variable) and the firms' creditworthiness into account. Firms pay a credit risk premium according to their creditworthiness that is modelled by a monotonically increasing function of their financial fragility. A firm's financial fragility is defined as the ratio between the outstanding debt and the liquid financial resources of the firm \citep{Delli-Gatti:2011aa}. Specifically the interest rate for firm $\kappa$, borrowing from bank $i$ is given by 
\begin{equation}
	r^\kappa_i(t) = \bar{r} (1+\chi_i(t) \mu (l_\kappa(t)) \quad, \label{firm_ir} 
\end{equation}
where $\bar{r}$ is a benchmark interest rate, $\chi_i(t)$ is the specificity of bank $i$ -- modelled as random variations in its operating costs, strategy, etc. and captured by a uniformly distributed random variable on the interval $(0, 1)$. $\mu (l_\kappa(t))$ is a proxy for the financial fragility of the borrower -- modelled by a monotonically increasing function $\mu(\cdot)$ of the borrower's debt to liquidity ratio $l_\kappa(t)$. The hyperbolic tangent is chosen for $\mu(\cdot)$. 

\subsection{The interbank market} \label{model_details_ib} Banks try to provide firm loans and grant them if they have enough liquidity. If they do not have enough cash, they approach other banks in the interbank market to obtain the required amount. If a bank does not have enough cash, and cannot raise the full amount for the requested firm loan on the interbank market, it does not pay out the loan. Interbank and firm loans have the same duration. Additional refinancing costs of banks remain with the firms. Each time step firms and banks repay $\tau$ percent of their outstanding debt. If banks have excess liquidity they offer it on the interbank market. The interbank market is modelled after an electronic marketplace where, in principle, all participants can enter into business relationships. In the model, banks choose the interbank offer with the most favourable rate. This does not mean that the emerging interbank network is fully connected. Emerging interbank networks are shown in \cref{network} and (weighted) degree distributions can be found in \cref{degree_dist}. Interbank rates $r_{ij}(t)$ offered by bank $i$ to bank $j$ take into account the specificity of bank $i$ and the creditworthiness of bank $j$. Specifically the interest rate on the interbank market for bank $j$ borrowing from bank $i$ is given by 
\begin{equation}
	r_{ij}(t) = \bar{r} (1+\psi_i(t) \mu (l_j(t)) \quad, \label{bank_ir} 
\end{equation}
where $\bar{r}$ is a benchmark interest rate, $\psi_i(t)$ is the specificity of bank $i$, modelled as random variations in its operating costs, strategy, etc. and captured by a uniformly distributed random variable on the interval $(0, 0.1)$. $\mu (l_j(t))$ is a proxy for the financial fragility of the borrower, modelled by a monotonically increasing function $\mu(\cdot)$ of the borrower's leverage $l_j(t)$. As the monotonically increasing function again the hyperbolic tangent is chosen. Banks add the additional refinancing costs to the offered interest rate for firms. Therefore the interest rate for firm $\kappa$, borrowing from bank $i$, which requires additional liquidity from bank $j$ is given by 
\begin{multline}
	r^\kappa_{ij}(t) = r^\kappa_i(t) + \frac{l_{jik}(t)}{b_\kappa(t)} r_{ji}(t) = \\
	= \bar{r} \left(1+\chi_i(t) \mu (l_\kappa(t)) + \frac{l_{jik}(t)}{b_\kappa(t)} \psi_j(t) \mu (l_i(t))\right)\quad, \label{firm_ir_ib} 
\end{multline}
where $b_\kappa(t)$ is the firm loan and $l_{jik}(t)/b_\kappa(t)$ is the ratio between the interbank and the firm loan. 

\subsection{Implementation of the systemic risk tax and the Tobin tax} A systemic risk premium, in form of the SRT, is imposed on all interbank transactions. Before entering a desired loan ${l}_{ijk}(t)$, the credit seeking banks $i$ can get quotes for the $SRT^{(+k)}_{ij}(t)$ rates from the central bank, for various banks $j$. They choose the interbank offer from bank $j$ with the smallest total rate, which is composed of $r^{\rm total}_{ij}(t) = r_{ij}(t)+ SRT^{(+k)}_{ij}(t)$. All other transactions are exempted from the SRT. In contrast to current market practice, the effective interest rate reflects both the creditworthiness of the borrowing counterparty and the SR increase associated with each transaction. The SRT is collected in a bailout fund. The SRT from the main text is given by 
\begin{multline}
	SRT_{ij}^{(+k)}(t) = \zeta \max \\
	\left[0, \sum_i p_i(t) \left(V^{(+k)}(t)R^{(+k)}_i(t) - V(t)R_i(t) \right) \right] \quad.  
\end{multline}
For $p_i(t)$ we use a proxy for the financial fragility of the borrower, modelled by a monotonically increasing function $p_i(t)=0.01\mu(l_i(t))$ of the borrower's leverage $l_i(t)$ at time $t$.

For comparison we implement a FTT (Tobin tax \citep{Tobin:1978aa}) for interbank loans. We impose a constant tax rate of 0.2\% of the transaction (this is about 5\% of the interbank interest rates) on all interbank rates on offer. Other transactions are not taxed. The FTT makes lending less attractive for firms that borrow from banks requiring liquidity from the interbank market, as refinancing costs remain with the firms.

Interbank rates $r_{ij}(t)$ offered by bank $i$ to bank $j$ including the FFT or the SRT are composed of 
\begin{equation}
	r^{\rm total}_{ij}(t) = r_{ij}(t)+ TAX \quad. 
\end{equation}
In case of the FFT the tax term is simply a constant tax rate of 0.2\% 
\begin{equation}
	r^{\rm total}_{ij}(t) = r_{ij}(t)+ 0.002 \quad. 
\end{equation}
To obtain a tax rate, the SRT must be expressed as a ratio with respect to the interbank loan ($SRT^{(+k)}_{ij}(t)/l_{jik}(t)$). The total rate is then given by 
\begin{equation}
	r^{\rm total}_{ij}(t) = r_{ij}(t)+ \frac{SRT_{ij}^{(+k)}(t)}{l_{ijk}(t)}\quad, 
\end{equation}
where $r_{ij}(t)$ is from \cref{bank_ir}.
Banks add the additional refinancing costs, including taxes, to the offered interest rate for firms. Therefore \cref{firm_ir} becomes for the Tobin-like tax 
\begin{multline}
	r^\kappa_{ij}(t) = r^\kappa_i(t) + \frac{l_{jik}(t)}{b_\kappa(t)} r^{\rm total}_{ji}(t) = \\
	= \bar{r} \left(1+\chi_i(t) \mu (l_\kappa(t)) + \frac{l_{jik}(t)}{b_\kappa(t)} \psi_j(t) \mu (l_i(t)) \right) + \\
	+ \frac{l_{jik}(t)}{b_\kappa(t)}0.002 \quad, \label{firm_ir_tax} 
\end{multline}
and in case of the SRT, 
\begin{multline}
	r^\kappa_{ij}(t) = r^\kappa_i(t) + \frac{l_{jik}(t)}{b_\kappa(t)} r^{\rm total}_{ji}(t) = \\
	= \bar{r} \left(1+\chi_i(t) \mu (l_\kappa(t)) + \frac{l_{jik}(t)}{b_\kappa(t)} \psi_j(t) \mu (l_i(t)) \right) + \\
	+ \frac{SRT_{ji}^{(+k)}(t)}{b_\kappa(t)} \quad. \label{firm_ir_srt} 
\end{multline}

\subsection{Model parameters} All parameters of the model are collected in \cref{parameters}. 
\begin{table*}
	\caption{List of the parameters as used in the model. \label{parameters}} 
	\begin{center}
		\begin{tabular}
			{ l l r } number of banks & $ B=20 $ \\
			number of firms/capitalists & $ F=100 $ \\
			number of workers (households) & $ H=1300 $ \\
			share of dividends & $ div=0.2 $ \\
			general refinancing rate & $ \bar{r}=0.02 $ \\
			labour productivity & $ \alpha=0.1 $ \\
			credit demand contraction & $ \phi=0.8 $ \\
			rate of debt reimbursement & $ \tau=0.05 $ \\
			wage rate & $ wb=1 $ \\
			number of applications in consumption goods market & $ z=2 $ \\
			propensity to consume & $ c=0.8 $ \\
			number of applications in credit market & $ n=5 $ \\
		\end{tabular}
	\end{center}
\end{table*}

\section{Comparison of different tax rates for the Tobin-like financial transaction tax} \label{tobin} 
\begin{figure*}
	\begin{center}
		\includegraphics[width=.99\columnwidth]{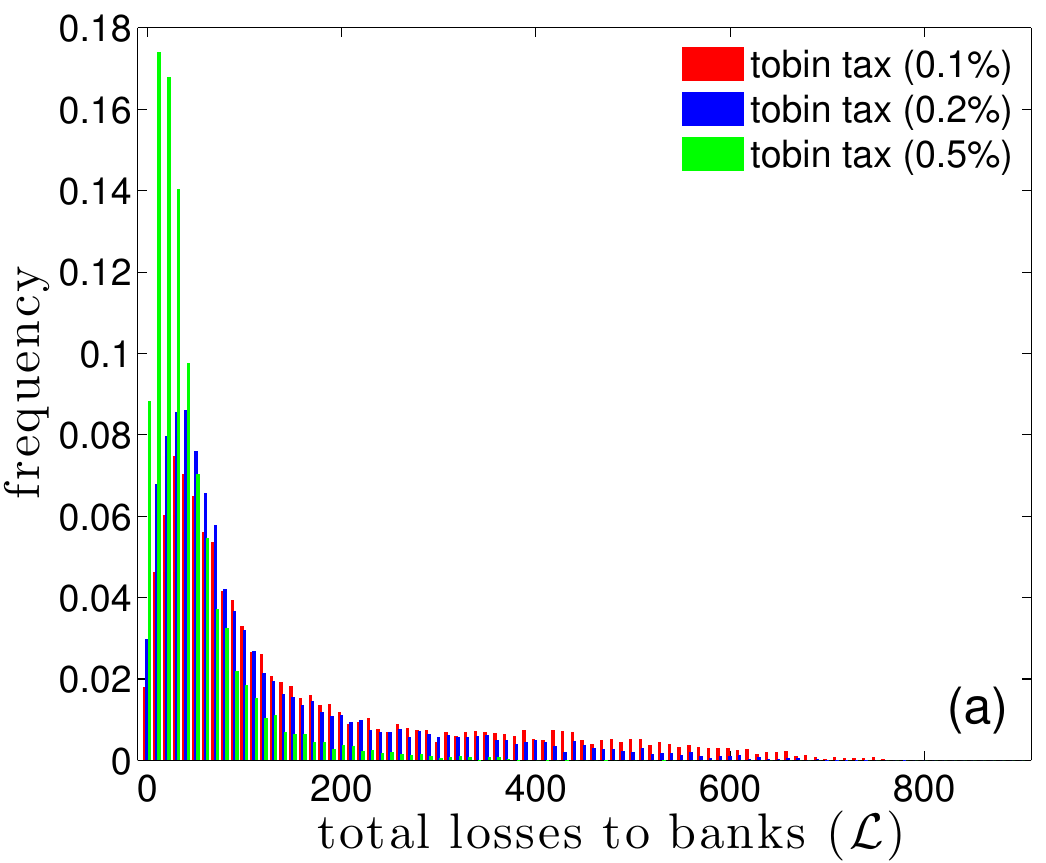} 
		\includegraphics[width=.99\columnwidth]{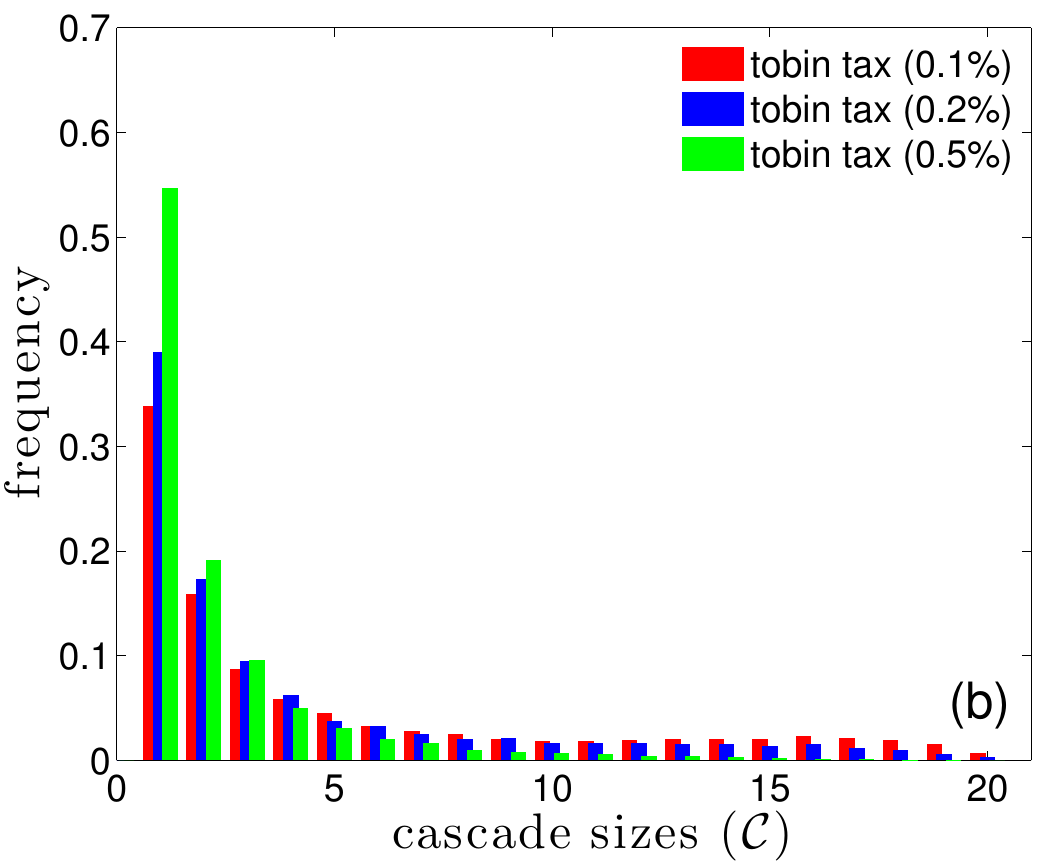} \\
		\includegraphics[width=.99\columnwidth]{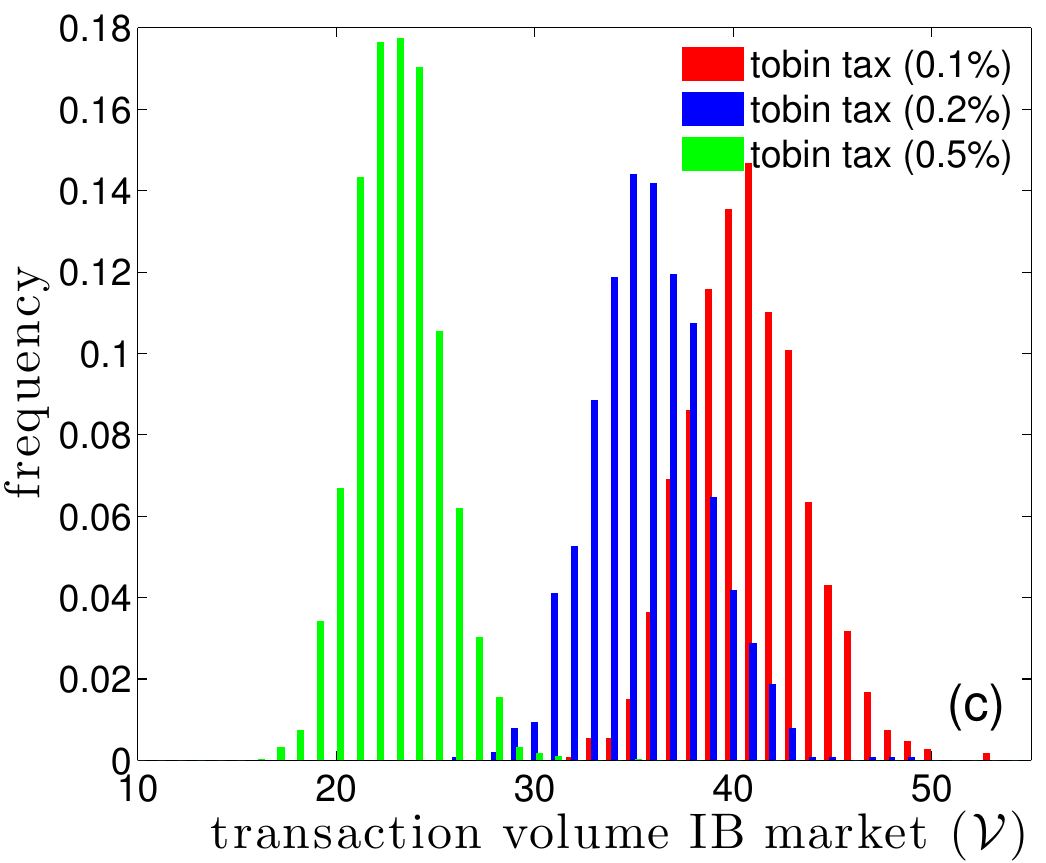} 
		\includegraphics[width=.99\columnwidth]{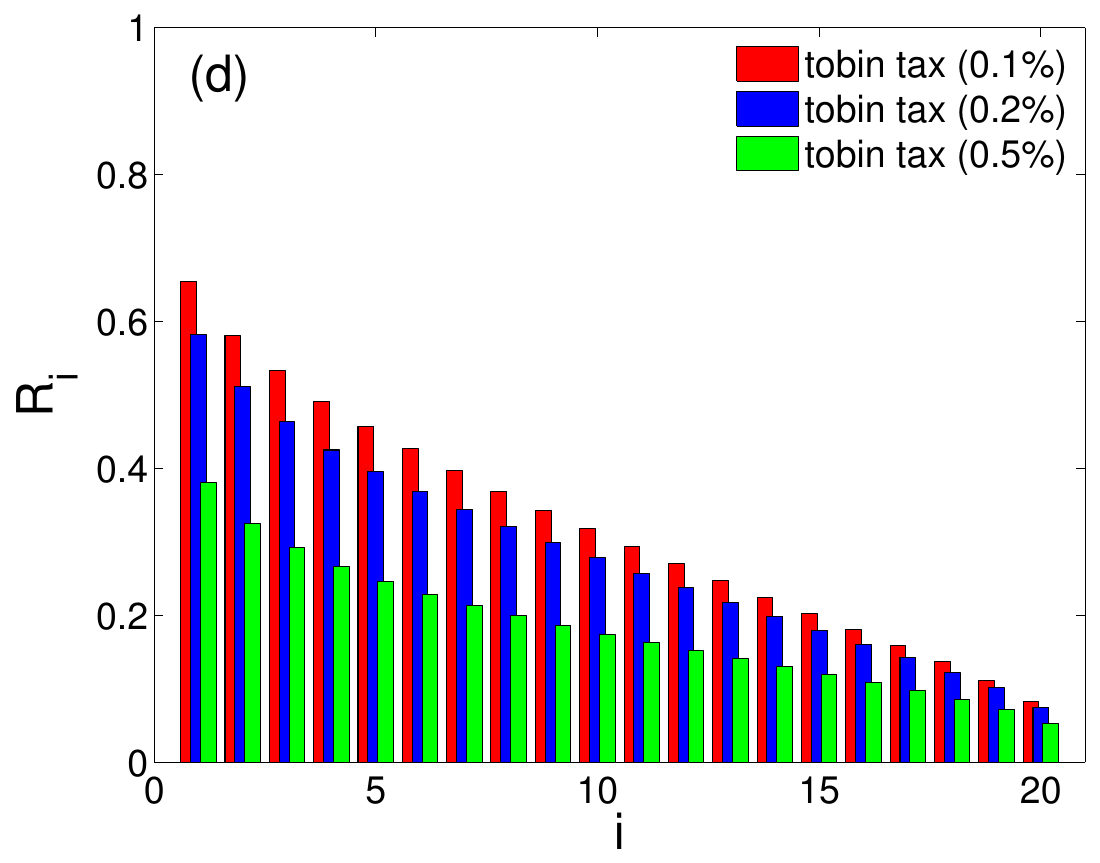} 
	\end{center}
	\caption{Comparison of different tax rates for the Tobin-like financial transaction tax, 0.1\% (red), 0.2\% (blue) and 0.5\% (green). (a) Distribution of total losses to banks ${\cal L}$, (b) distribution of cascade sizes ${\cal C}$ of defaulting banks and (c) distribution of total transaction volume in the interbank market ${\cal V}$, (d) distribution of DebtRank $R_{i}$. Banks are ordered by DebtRank, the most important being to the very left. $10,000$ independent, identical simulations, each with $500$ time steps, $20$ banks.} \label{hist_tobin} 
\end{figure*}
In \cref{hist_tobin} we show the distribution functions of the three measures for (a) losses ${\cal L}$, (b) cascade sizes ${\cal C}$, (c) transaction volume in the interbank market ${\cal V}$ and (d) the distribution of DebtRank $R_{i}$, for the simulations performed with different tax rates for the Tobin-like FTT, 0.1\% (red), 0.2\% (blue) and 0.5\% (green). Clearly, the shape of the distribution of losses ${\cal L}$ and cascade sizes ${\cal C}$ are similar. The tail of the distributions is only reduced due to a decrease in efficiency (transaction volume), as can be seen in \cref{hist_tobin}(c). Evidently, average losses ${\cal L}$ are reduced at the cost of a loss of efficiency by roughly the same factor.

\begin{figure*}
	\begin{center}
		\includegraphics[width=.99\columnwidth]{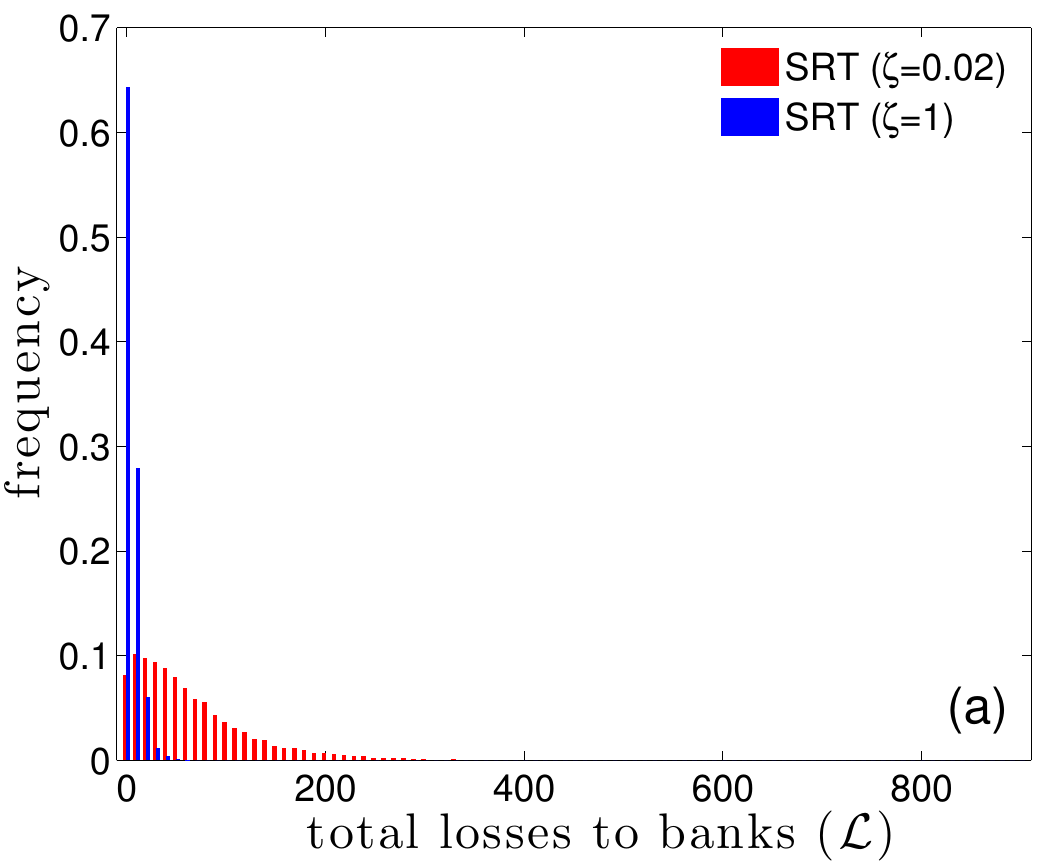} 
		\includegraphics[width=.99\columnwidth]{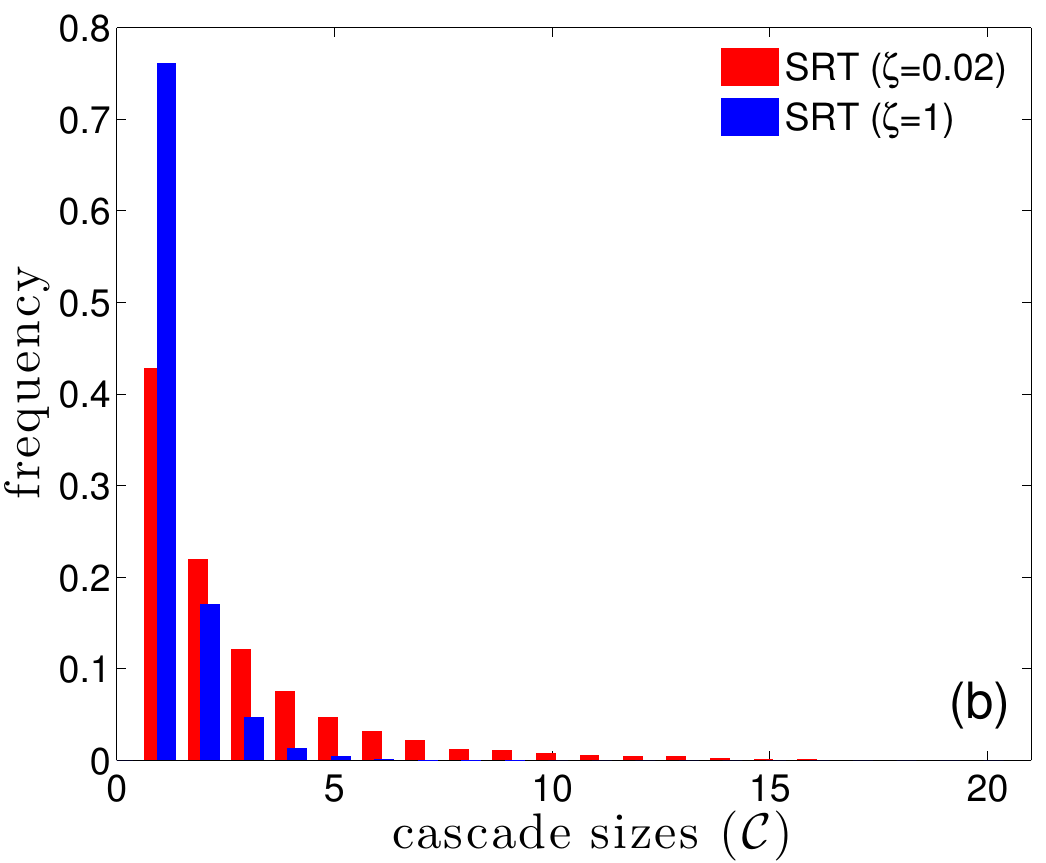} \\
		\includegraphics[width=.99\columnwidth]{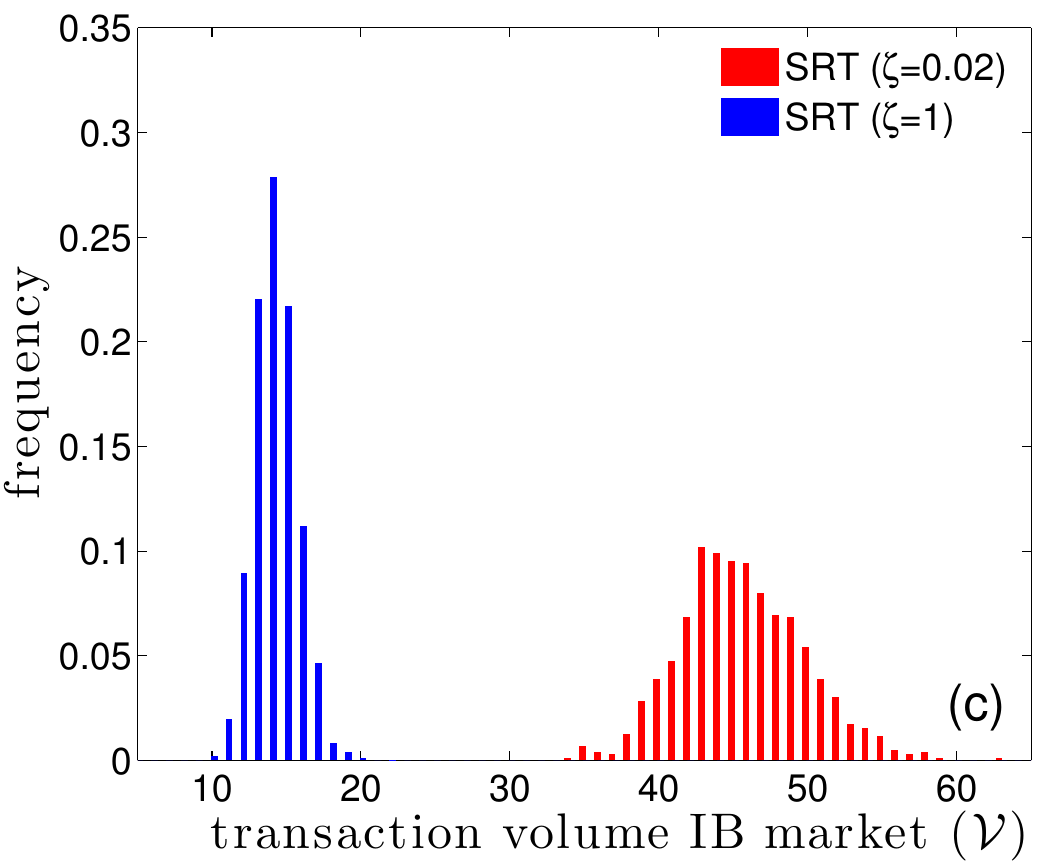} 
		\includegraphics[width=.99\columnwidth]{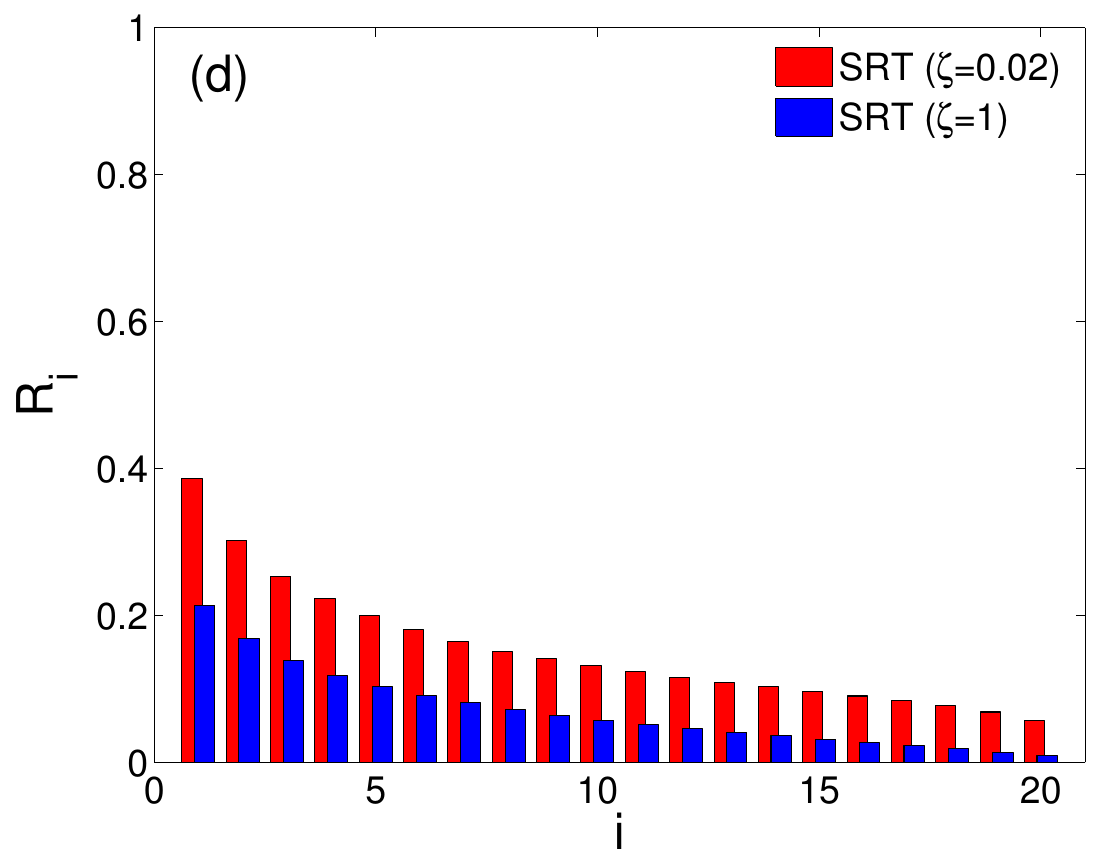} 		
	\end{center}
	\caption{Comparison of different levels of the systemic risk tax, $\zeta=0.02$ (red) and $\zeta=1$ (blue). (a) Distribution of total losses to banks ${\cal L}$, (b) distribution of cascade sizes ${\cal C}$ of defaulting banks and (c) distribution of total transaction volume in the interbank market ${\cal V}$, (d) distribution of DebtRank $R_{i}$. Banks are ordered by DebtRank, the most important being to the very left. $10,000$ independent, identical simulations, each with $500$ time steps, $20$ banks.} \label{hist_srt} 
\end{figure*}
For the comparison of different levels of the SRT we choose $\zeta=0.02$ (red) and $\zeta=1$ (blue), as shown in \cref{hist_srt}. Again, we compare the three measures for (a) losses ${\cal L}$, (b) cascade sizes ${\cal C}$, (c) transaction volume in the interbank market ${\cal V}$ and (d) the distribution of DebtRank $R_{i}$. Clearly, for both $\zeta$ the SRT gets completely rid of big losses in the system. $\zeta=1$ reduces average losses ${\cal L}$ by a factor of $2$ compared to the case of $\zeta=0.02$, at the cost of a loss of efficiency by roughly the same factor, as can be seen in \cref{hist_srt}(c). The SRT with ($\zeta=1$) leads to homogeneous SR spreading across all agents, as shown in \cref{hist_srt}(d). 
\begin{figure}
	\begin{center}
		\includegraphics[width=.99\columnwidth]{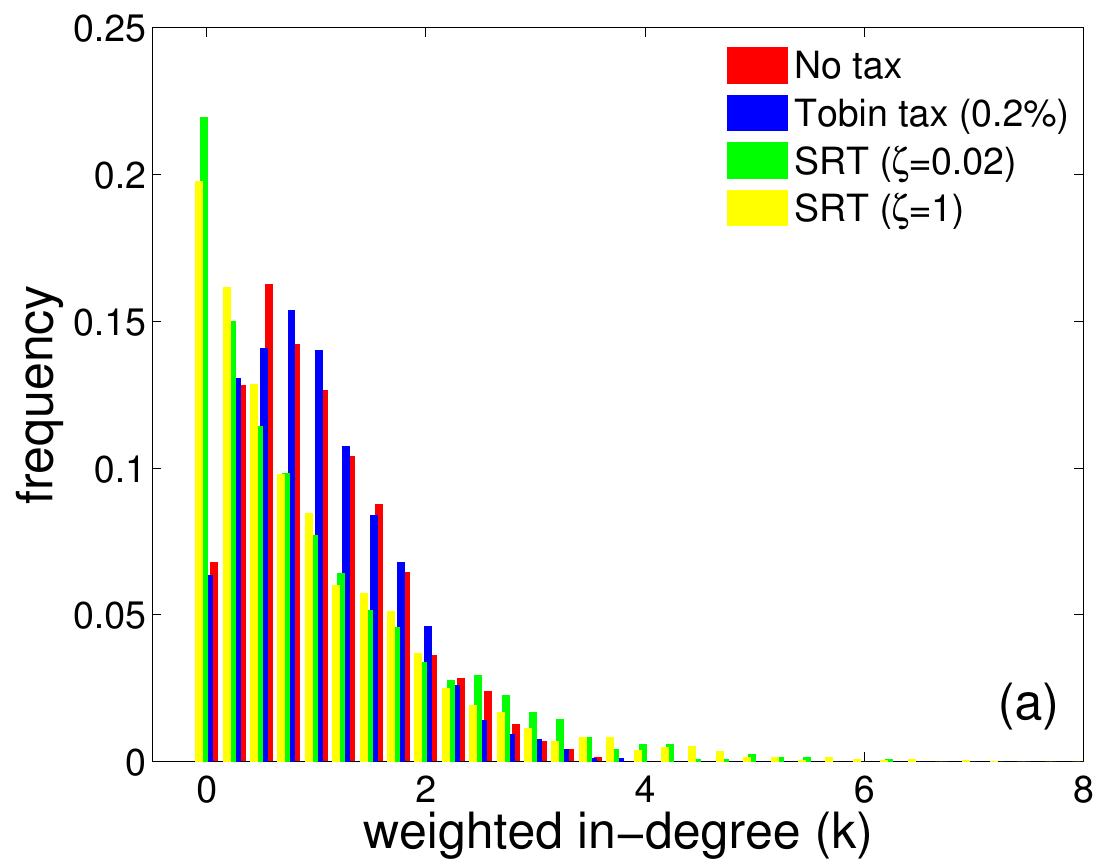} 
		\includegraphics[width=.99\columnwidth]{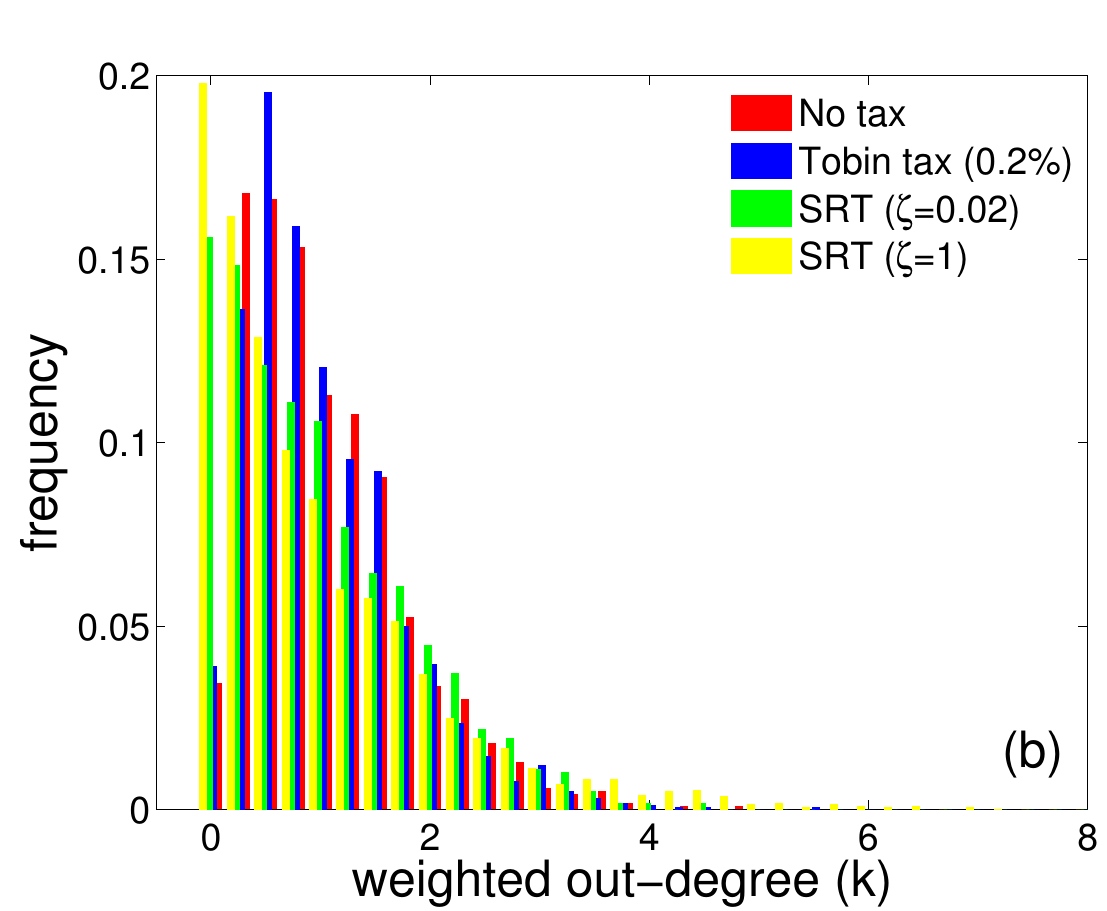} 
		\includegraphics[width=.99\columnwidth]{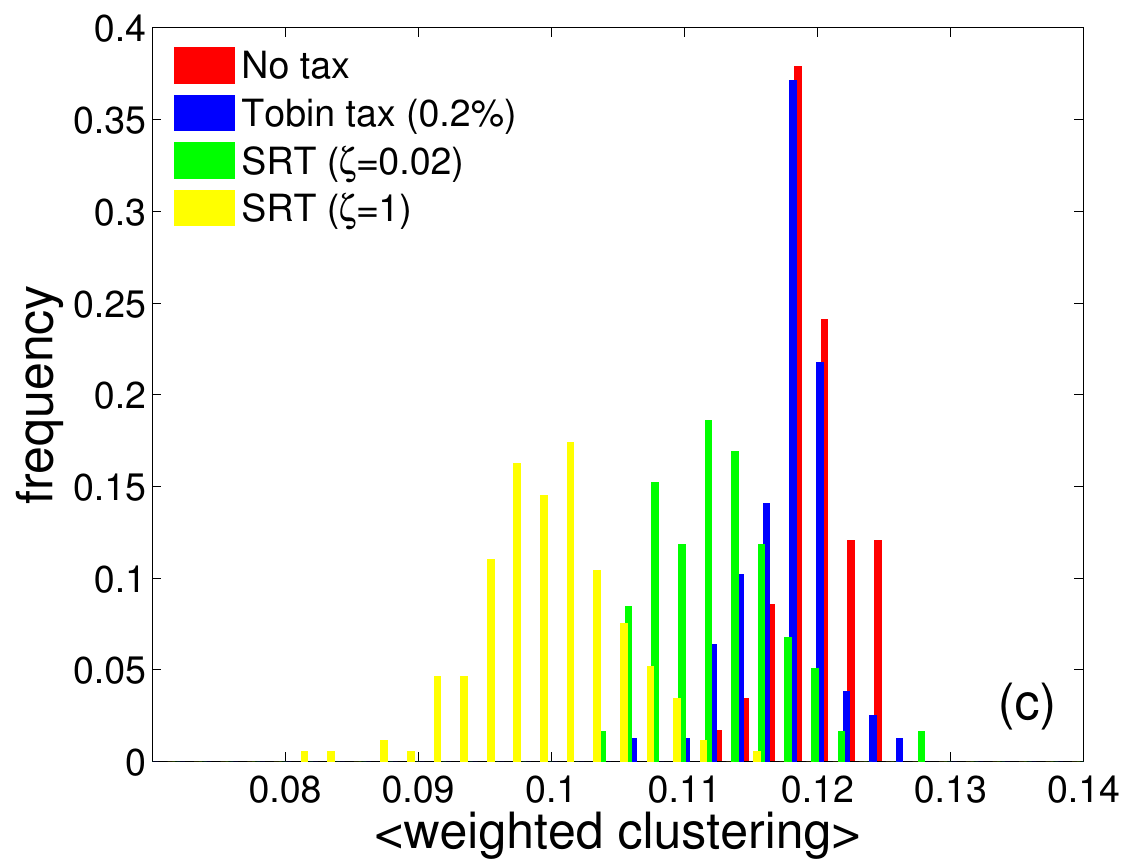} 
	\end{center}
	\caption{The effect of the bank selection process induced by the systemic risk tax on the interbank liability network topology. Distributions of weighted in-degrees $k$ (a), and weighted out-degrees $k$ (b) of the interbank liability network $(L^{\rm net}_{ij}(t))$, without FTT (red), 0.2\% Tobin tax (blue), the SRT ($\zeta=0.02$) (green) and the SRT ($\zeta=1$) (yellow). (c) Shows the average weighted clustering coefficient of the interbank liability network $(L^{\rm net}_{ij}(t))$ from the model. The weighted in-degrees distributions are clearly affected in the SRT modes. The weighted out-degree distribution is mainly influenced by the cash needs of a bank. Therefore, the weighted out-degree distribution of the SRT modes is less clearly affected. The distributions are from an average over $1000$ simulations runs and show the situation at time $t = 100$.} \label{degree_dist} 
\end{figure}
\begin{table*}
	\caption{Network measures \label{nwanalysis}} 
	\begin{center}
		\begin{tabular}
			{ l l l l l l l l l l l l l l l l l} Mode & $\langle k \rangle$ & $\langle k^{weighted} \rangle$ & $\langle C_i \rangle$ & $\langle C_i^{weighted} \rangle$ & $\langle g_i \rangle$ & $\langle g^{weighted}_i \rangle$\\
			\hline Normal & 9.43(5) & 38.4(35) & 0.136(5) & 0.119(3) & 10.3(7) & 40.2(53) \\
			Tobin tax & 9.39(9) & 32.4(32) & 0.135(5) & 0.117(3) & 10.4(7) & 40.0(55) \\
			SRT ($\zeta=0.02$) & 9.15(13) & 25.4(35) & 0.138(6) & 0.112(5) & 11.9(20) & 44.1(76) \\
			SRT ($\zeta=1$) & 8.73(20) & 7.4(14) & 0.122(6) & 0.100(4) & 11.3(12) & 45.9(68) \\
		\end{tabular}
	\end{center}
\end{table*}
In \cref{degree_dist} we show the effect of the bank selection process induced by the SRT on the interbank liability network topology. The distributions of weighted in-degrees $k$ of the interbank liability network $(L^{\rm net}_{ij}(t))$, without FTT (red), 0.2\% Tobin tax (blue), the SRT ($\zeta=0.02$) (green) and, the SRT ($\zeta=1$) (yellow) are shown in \cref{degree_dist}(a). Without FTT, the emerging liability network shows Poisson distributed in-degrees. The interbank network topology without FTT coincides nicely with the expected result from random linking. In the SRT modes, market participants looking for credit will try to avoid the tax by looking for credit opportunities that do not increase SR and are thus tax-free. This leads to fewer banks lending on the interbank market and is reflected in \cref{degree_dist}(a) by the high number of nodes with a low weighted in-degree.
 
The total demand for interbank loans (which is approximately the same for the SRT with $\zeta=0.02$ as without FTT) is now serviced by fewer banks. As a result, the in-degree distribution of the SRT mode broadens and has a fat tail. The out-degree distribution is mainly influenced by the cash needs of a bank. Therefore, the weighted out-degree distribution of the SRT modes is less clearly affected, which is shown in \cref{degree_dist}(b).

In \cref{degree_dist}(c) we show the average weighted clustering coefficient of the interbank liability network $(L^{\rm net}_{ij}(t))$, without FTT (red), 0.2\% Tobin tax (blue), the SRT ($\zeta=0.02$) (green) and the SRT ($\zeta=1$) (yellow). Average weighted clustering coefficients are calculated according to \citet{Barrat:2004aa}. The average clustering is roughly the same without a FTT on interbank loans and for the 0.2\% Tobin tax. The SRT reduces average clustering, as can be seen in \cref{degree_dist}(c). 

Mean values of various centrality measures, averaged over $1000$ simulations runs, can be found in \cref{nwanalysis}. $\langle k \rangle$ and $\langle k^{weighted} \rangle$ shows the mean degree and the weighted mean degree for the different modes. $\langle k \rangle$ is approximately the same for all modes. $\langle k^{weighted} \rangle$ shows the largest value for the normal mode and lower values in all other modes. Clearly, with the SRT ($\zeta=1$)  $\langle k^{weighted} \rangle$ is substantially reduced. With $\langle C_i \rangle$ and $\langle C_i^{weighted} \rangle$ we show values for the average clustering coefficients and the average weighted clustering coefficients from \cref{degree_dist}(c). Additionally, we provide values for the average betweenness centrality ($\langle g_i \rangle$) and the average weighted betweenness centrality ($\langle g^{weighted}_i \rangle$) for the different modes. The SRT increases both the $\langle g_i \rangle$ and the $\langle g^{weighted}_i \rangle$.

\section{Empirical data} \label{data}
Data provided by the Austrian Central Bank (OeNB) contains fully anonymised, and linearly transformed interbank liabilities/exposures $L^{\rm data}_{ij}(t)$ from the entire Austrian banking system, comprised of about 800 banks over 12 consecutive quarters from 2006-2008. The data set additionally includes total assets, total liabilities, assets due from banks, liabilities due to banks, and liquid assets (without interbank assets/liabilities) for all banks again in anonymised form. The data does not contain credit ratings of banks. Therefore we assume $p_i= 0.0025$ for all banks in the data set. This corresponds approximately to Standard \& Poor's One-Year Global Corporate Default Rates for Rating Categories A+, A, and BBB+ in 2008 \citep{standardandpoors}. Representative Austrian banks are in the Rating Categories A+, A and A-.

\section{DebtRank} \label{debtrank_section} 
DebtRank is a recursive method suggested in \citet{Battiston:2012aa} to determine the systemic importance of nodes in financial networks. It is a number measuring the fraction of the total economic value in the network that is potentially affected by a node or a set of nodes. $L_{ij}$ denotes the interbank liability network at any given moment (loans of bank $j$ to bank $i$), and $C_{i}$ is the capital of bank $i$. If bank $i$ defaults and cannot repay its loans, bank $j$ loses the loans $L_{ij}$. If $j$ does not have enough capital available to cover the loss, $j$ also defaults. The impact of bank $i$ on bank $j$ (in case of a default of $i$) is therefore defined as 
\begin{equation}
	\label{impact} W_{ij} = \min \left[1,\frac{L_{ij}}{C_{j}} \right] \quad. 
\end{equation}
The value of the impact of bank $i$ on its neighbours is $I_{i} = \sum_{j} W_{ij} v_{j}$. The impact is measured by the {\em economic value} $v_{i}$ of bank $i$. For the economic value we use two different proxies. Given the total outstanding interbank liabilities of bank $i$, $L_{i}=\sum_{j}L_{ji}$, its economic value is defined as 
\begin{equation}
	\label{ecovalue} v_{i}=L_{i}/\sum_{j}L_{j} \quad. 
\end{equation}
To take into account the impact of nodes at distance two and higher, this has to be computed recursively. If the network $W_{ij}$ contains cycles, the impact can exceed one. To avoid this problem an alternative was suggested in \citet{Battiston:2012aa}, where two state variables, $h_{\rm i}(t)$ and $s_{\rm i}(t)$, are assigned to each node. $h_{\rm i}$ is a continuous variable between zero and one; $s_{\rm i}$ is a discrete state variable for three possible states, undistressed, distressed, and inactive, $s_{\rm i} \in \{U, D, I\}$. The initial conditions are $h_{i}(1) = \Psi \, , \forall i \in S ;\; h_{i}(1)=0 \, , \forall i \not \in S$, and $s_{i}(1) = D \, , \forall i \in S ;\; s_{i}(1) = U \, , \forall i \not \in S$ (parameter $\Psi$ quantifies the initial level of distress: $\Psi \in [0, 1]$, with $\Psi = 1$ meaning default). The dynamics of $h_i$ is then specified by 
\begin{equation}
	h_{i}(t) = \min\left[1,h_{i}(t-1)+\sum_{j\mid s_{j}(t-1) = D} W_{ ji}h_{j}(t-1) \right] \quad. 
\end{equation}
The sum extends over these $j$, for which $s_{j}(t-1) = D$, 
\begin{equation}
	s_{i}(t) = 
	\begin{cases}
		D & \text{if } h_{i}(t) > 0; s_{i}(t-1) \neq I ,\\
		I & \text{if } s_{i}(t-1) = D , \\
		s_{i}(t-1) & \text{otherwise} \quad. 
	\end{cases}
\end{equation}
The DebtRank of the set $S$ (set of nodes in distress at time $1$), is $R^{\prime}_S = \sum_{j} h_{j}(T)v_{j} - \sum_{j} h_{j}(1)v_{j}$, and measures the distress in the system, excluding the initial distress. If $S$ is a single node, the DebtRank measures its systemic importance on the network. The DebtRank of $S$ containing only the single node $i$ is 
\begin{equation}
	\label{debtrank} R^{\prime}_{i} = \sum_{j} h_{j}(T)v_{j} - h_{i}(1)v_{i} \quad. 
\end{equation}
The DebtRank, as defined in \cref{debtrank}, excludes the loss generated directly by the default of the node itself and measures only the impact on the rest of the system through default contagion. For some purposes, however, it is useful to include the direct loss of a default of $i$ as well. The total loss caused by the set of nodes $S$ in distress at time $1$, including the initial distress is 
\begin{equation}
	\label{debtrank_self} R_S = \sum_{j} h_{j}(T)v_{j} \quad. 
\end{equation}

\section{Derivation of the expected systemic loss} \label{el_approx}
To compute the expected systemic loss, we first consider the simple case where only one bank $i$ can default and all other $b-1$ banks survive.  In this case the expected loss is given by ${\rm EL}_{i}^{\rm syst}({\rm one\,\, default}) = V \cdot p_i \cdot (1-p_1) \cdot \ldots \cdot (1-p_{i-1}) \cdot (1-p_{i+1}) \cdot \ldots \cdot (1-p_{b}) \cdot R_i$, where $p_i$ is the probability of default of bank $i$, and $(1-p_j)$ the survival probability of $j$. The general case occurs when we also consider possible joint defaults, meaning that a set of banks $S$ go into distress. Taking into account {\em all} possible combinations of defaulting and surviving banks, we arrive at a combinatorial expression of the expected loss for an economy of $b$ banks  
\begin{equation}
	{\rm EL}^{\rm syst} = V \sum_{S \in \mathcal{P}(B)} \prod_{i \in S} p_i \prod_{j \in B \setminus S} (1-p_j) \: R_S \quad, \label{totEL_normalized} 
\end{equation}
where $\mathcal{P}(B)$ is the power set of the set of banks $B$, and $R_S$ is the DebtRank of the set $S$ of nodes initially in distress. $R_{\emptyset}$, the DebtRank of the empty set is defined as zero. The reason is that, by definition of DebtRank, $R_S \leq 1$, the value obtained in \cref{totEL_normalized} cannot exceed the total economic value. 

\Cref{totEL_normalized} is only practical for situations with less than about $20-30$ banks. Computing the power set and calculating DebtRanks for all possible combinations of more than $30$ banks in a large financial networks is practically unfeasible. If the default probabilities are low ($p_i \ll 1$) or the interconnectedness is low ($R_i \approx v_i$), $R_S$ can be approximated by
\begin{equation}
	R_S \approx \sum_{i \in S} R_i \quad. \label{debtrank_approx} 
\end{equation}
In an unconnected or unleveraged financial system ($R_i = v_i$), $R_S$ is exactly equal to $\sum_{i \in S} R_i$. If $p_i \ll 1$, the first terms of \cref{totEL_normalized} (with only one node initially in distress) contribute more to the final result. Thus the approximation \cref{debtrank_approx} has only a minor impact on the final result. Typically, $p_i \ll 1$ or $R_i \approx v_i$ holds in real word financial networks.
With the approximation \cref{debtrank_approx}, \cref{EL} can be derived from \cref{totEL_normalized} by
\begin{align}
	& {\rm EL}^{\rm syst} \approx V \sum_{S \in \mathcal{P}(B)} \prod_{i \in S} p_i \prod_{j \in B \setminus S} (1-p_j) \left(\sum_{i \in S} R_i \right) \\
	& = V \sum_{i=1}^{B}\underbrace{\left(\sum_{J \in \mathcal{P}(B \setminus \{i\})} \prod_{j \in J} p_j \prod_{k \in B \setminus (J \cup \{i\})} (1-p_k) \right)}_{=1}  p_i \, R_i \label{brace} \\
	& = V \sum_{i=1}^{B} p_i \, R_i \quad.
	\label{totEL_approx} 
\end{align}
The term in brackets in \cref{brace} sums to 1 (proof by induction). This approximation is practical for large financial networks, for details see \citep{Poledna:2015aa}. 
 
\section{Measures for losses, default cascades and transaction volume} \label{risk_measures} We use the following three observables: (1) the size of the cascade, ${\cal C}$ as the number of defaulting banks triggered by an initial bank default ($1\leq {\cal C} \leq B$), (2) the total losses to banks following a default or cascade of defaults, ${\cal L}= \sum_{i \in I}\sum_{j=1}^B L_{ ij}(t)$, where $I$ is the set of defaulting banks, and (3) the average transaction volume in the interbank market in simulation runs longer than $100$ time steps, 
\begin{equation}
	\mathcal{V}=\frac{1}{T}\sum_{t=1}^T\sum_{j=1}^B\sum_{i=1}^B \sum_{k \in K}l_{jik}(t) \quad, 
\end{equation}
where $K$ represents new interbank loans at time step $t$.

\end{document}